\documentclass{aastex7}

\def\Msun{M_\odot}

\newcommand{\hii}{H{\sc i}~21\,cm}
\newcommand{\hi}{H{\sc i}}

\newcommand{\kms}{km~s$^{-1}$}

\usepackage{multirow}
\usepackage{float}
\usepackage{needspace}
\begin{document}

\title{The atomic gas properties of Green Pea galaxies: Connections to Lyman continuum leakage}

\author{Archishman Khasnovis}
\affiliation{National Centre for Radio Astrophysics, Tata Institute of Fundamental Research, Pune University, Pune-411007, India}
\email{archishman@ncra.tifr.res.in}

\author[0000-0002-9757-7206]{Nissim Kanekar}
\affiliation{National Centre for Radio Astrophysics, Tata Institute of Fundamental Research, Pune University, Pune-411007, India}
\email{nkanekar@ncra.tifr.res.in}

\author[0000-0002-9226-5350]{Sangeeta Malhotra}
\affiliation{Astrophysics Division, NASA Goddard Space Flight Center, Greenbelt, MD 20771, USA}
\email{sangeeta.malhotra@nasa.gov}
\author{James Rhoads}
\affiliation{Astrophysics Division, NASA Goddard Space Flight Center, Greenbelt, MD 20771, USA}
\email{james.e.rhoads@nasa.gov}

\begin{abstract}

We have used the Green Bank Telescope to search for H{\sc{i}} 21\,cm emission from 30 Green Pea galaxies (GPs) at $z\approx0.012-0.045$, obtaining 7 detections of H{\sc{i}} 21\,cm emission and 17 upper limits on the H{\sc{i}} mass. Including GPs from the literature, we obtain a sample of 60 GPs at $z<0.05$, with 19 detections and 41 non-detections of H{\sc{i}} 21\,cm emission, and with stellar masses in the range $10^6-10^9\,\rm{M_{\odot}}$. We use the line luminosity ratio O32~$\equiv$~[O{\sc iii}]$\lambda 5007+\lambda 4959$/[O{\sc ii}]$\lambda$3727,3729 as an indicator of Lyman continuum (LyC) leakage, and examine the dependence of the H{\sc{i}} properties of the 60 GPs on the O32 ratio. We obtain a far higher H{\sc{i}} 21\,cm detection rate ($\approx53^{+16}_{-13}$\%) for the 32 GPs with O32~$<10$ than that ($7.1^{+9.4}_{-4.6}$\%) for the 28 GPs with O32~$>10$. We find statistically significant evidence that the H{\sc{i}} mass, the H{\sc{i}}-to-stellar mass ratio, and the H{\sc{i}} gas depletion timescale of GPs with O32~$>10$ are lower than the corresponding values for GPs with O32~$<10$. Earlier studies have shown that galaxies with O32~$>10$ tend to show significant LyC leakage: our results indicate that this is due to the lack of H{\sc{i}} in such galaxies, with most of the H{\sc{i}} consumed in the starburst. Our results further suggest that H{\sc{i}} 21\,cm studies of the galaxies that reionized the Universe at $z\gtrsim6$ are likely to find an anti-correlation between the H{\sc{i}} 21\,cm and Ly$\alpha$ emission signals, due to the paucity of H{\sc{i}} in the strongest LyC and Ly$\alpha$ leakers.
\end{abstract}

\keywords{Galaxies --- starburst, Galaxies --- dwarf, Galaxies --- 21cm line emission}

\section{Introduction} \label{sec:intro}

Green Pea galaxies (GPs) are dwarf starburst galaxies in the nearby Universe (at $z \lesssim 0.3$) that show strong [O{\sc iii}]$\lambda$5007\AA\ line emission which dominates the stellar continuum \citep{cardamone09}. These galaxies are interesting because they have properties similar to those of the high-redshift star-forming galaxies that are believed to have reionized the Universe at $z \gtrsim 6$ \citep[e.g.][]{fan06}. For example, GPs have low metallicity, low dust content, and compact or interacting morphology, and show intense starburst activity \citep[e.g.][]{malhotra12,jiang19}. They typically show strong Ly$\alpha$ emission \citep[e.g.][]{henry15,yang16,jaskot19,izotov20}, with a Ly$\alpha$ equivalent width distribution similar to that of Ly$\alpha$ emitters at $z \gtrsim 3$ \citep{yang16}, and their UV continuum sizes are similar to those of Ly$\alpha$ emitters at $z \approx 3-6$ \citep{yang17}. GPs also show high values of the extinction-corrected line luminosity ratio O32~$\equiv$~[O{\sc iii}]$\lambda 5007 + \lambda 4959$/[O{\sc ii}]$\lambda$3727,3729, suggesting the presence of ionized regions in the galaxies, through which Lyman-continuum (LyC) photons might escape \citep[e.g.][]{jaskot13,jaskot24}. Indeed, galaxies with LyC leakage at both low and high redshifts have been typically found to show high O32 ratios, $\gtrsim 10$ \citep[e.g.][]{nakajima14,flury22}. Most interestingly, a number of GPs at $z \approx 0.3$ have been found to show LyC leakage, with LyC escape fractions in the range $\approx 2.5 - 72$\% \citep[e.g.][]{izotov16, izotov17, izotov18}. This is similar to the expected LyC leakage from the galaxies that reionized the Universe \citep[e.g.][]{rhoads23,llerena24}. 

The process by which LyC photons escaped star-forming galaxies at $z \gtrsim 6$ and reionized the Universe is still not well understood, as these dwarf galaxies are faint and hard to study directly. Low-redshift analogs of these early galaxies, such as the GPs, offer the possibility of detailed studies to understand the process of LyC and Ly$\alpha$ leakage. 

Although the stellar, nebular, and star-formation properties of GPs have been well characterized by optical/UV studies, we still know little about the neutral atomic and molecular gas properties of these galaxies. Neutral atomic hydrogen (\hi) is critical to fuel the observed starburst activity in GPs, but the presence of significant amounts of \hi\ in these galaxies would also hinder the escape of LyC and Ly$\alpha$ photons. Connecting the \hi\ content and distribution in GPs to the star-formation and stellar properties is hence of much interest. Very recently, the \hi\ content of a sample of low-$z$ GPs has been measured for the first time \citep{kanekar21,chandola24}. This has been followed by \hii\ mapping of the \hi\ spatial distribution in two GPs at $z \approx 0.04$, yielding evidence of a major merger in both cases \citep{purkayastha22,purkayastha24}. 

At present, there are only $\approx 20$ detections of \hii\ emission from GPs \citep{kanekar21,chandola24}. Increasing the number of such \hii\ detections is critical, to connect the \hi\ properties of GPs to their stellar and nebular properties, and to enable follow-up \hii\ mapping studies to probe the \hi\ spatial distribution. Here, we report results from a  new Green Bank Telescope (GBT) \hii\ emission survey of a sample of GPs at $z \lesssim 0.045$.\footnote{Throughout the paper, we assume a flat $\Lambda$-cold dark matter ($\Lambda$CDM)  cosmology, with $\rm H_{0}=67.4$ $\rm km$ $\rm s^{-1} Mpc^{-1}$, $\Omega_{\Lambda}=0.685$ and $\Omega_{m}=0.315$ \citep{planck20}.} We further compare the \hi\ properties of a sample of GPs, drawn from our survey and the literature, to their O32 values, to obtain the first connections between \hi\ content and LyC leakage in these enigmatic galaxies.

\section{The GBT Observations, Data Analysis, and Results} 
\label{sec:obs}

We used the GBT L-band receivers to search for \hii\ emission from 30 GPs at $z < 0.045$ (proposal ID: GBT/22B-346; PI: N. Kanekar), with the Versatile GBT Astronomical Spectrometer (VEGAS) as the backend. The target GPs were selected from the large GP sample of \citet{jiang19}, based on Sloan Digital Sky Survey (SDSS) spectroscopy. We chose to restrict the sample to $z < 0.045$ to ensure a high \hi\ mass sensitivity, lower levels of radio frequency interference, and the possibility of follow-up \hii\ mapping studies at high angular resolution. The redshifted \hii\ line frequencies lie in the range $\approx 1360-1406$~MHz.

A bandwidth of 23.44~MHz was used for all sources, centred at the redshifted \hii\ line frequency, and sub-divided into 16,384 channels. This yielded a velocity coverage of $\approx \pm 2500$~\kms\ around the expected \hii\ line frequency and a velocity resolution of $\approx 0.6$~\kms, after Hanning smoothing and resampling. Position switching was used to calibrate the system bandpass, with On and Off scans each of 5m duration, and 2s integrations. The total observing time for the programme was $\approx 193$~hours, with on-source observing times of $\approx 0.5-5$~hours on individual sources. 

All data were analysed using the standard package {\sc gbtidl} \citep{gbtidl06}. For each source, each On/Off pair was initially calibrated separately for each polarization, and the calibrated spectra then shifted to the barycentric frame. Next, each calibrated spectrum was inspected for the presence of radio frequency interference (RFI) or other systematic effects between $-400$~\kms\ and $+400$~\kms\ of the expected redshifted \hii\ line frequency. Any spectra that were found to be affected by RFI or systematic non-Gaussian issues were removed. Following this, the ``clean'' spectra of each source were averaged together, to produce a single \hii\ spectrum for the source. Each source spectrum was then smoothed to, and resampled at, a velocity resolution of $\approx 10$~\kms, and inspected for the presence of line emission. Finally, a second-order polynomial was fitted to line-free channels in the above velocity range and subtracted out to obtain the final \hii\ line spectrum for the source. 
The final RMS noise values for all sources lie in the range $1.2-3.0$~mJy per 0.6~\kms\ channel.

The GBT observations of 6 of the target GPs were affected by RFI (especially at frequencies $\approx 1381$~MHz, due to transmissions from GPS satellites), rendering the data unusable. From the remaining 24 GPs, we obtained 7 clear detections (at $\gtrsim 7 \sigma$ significance) and 17 non-detections of \hii\  emission. 

\begin{figure}[t!]
\centering
\includegraphics[width=0.32\textwidth]{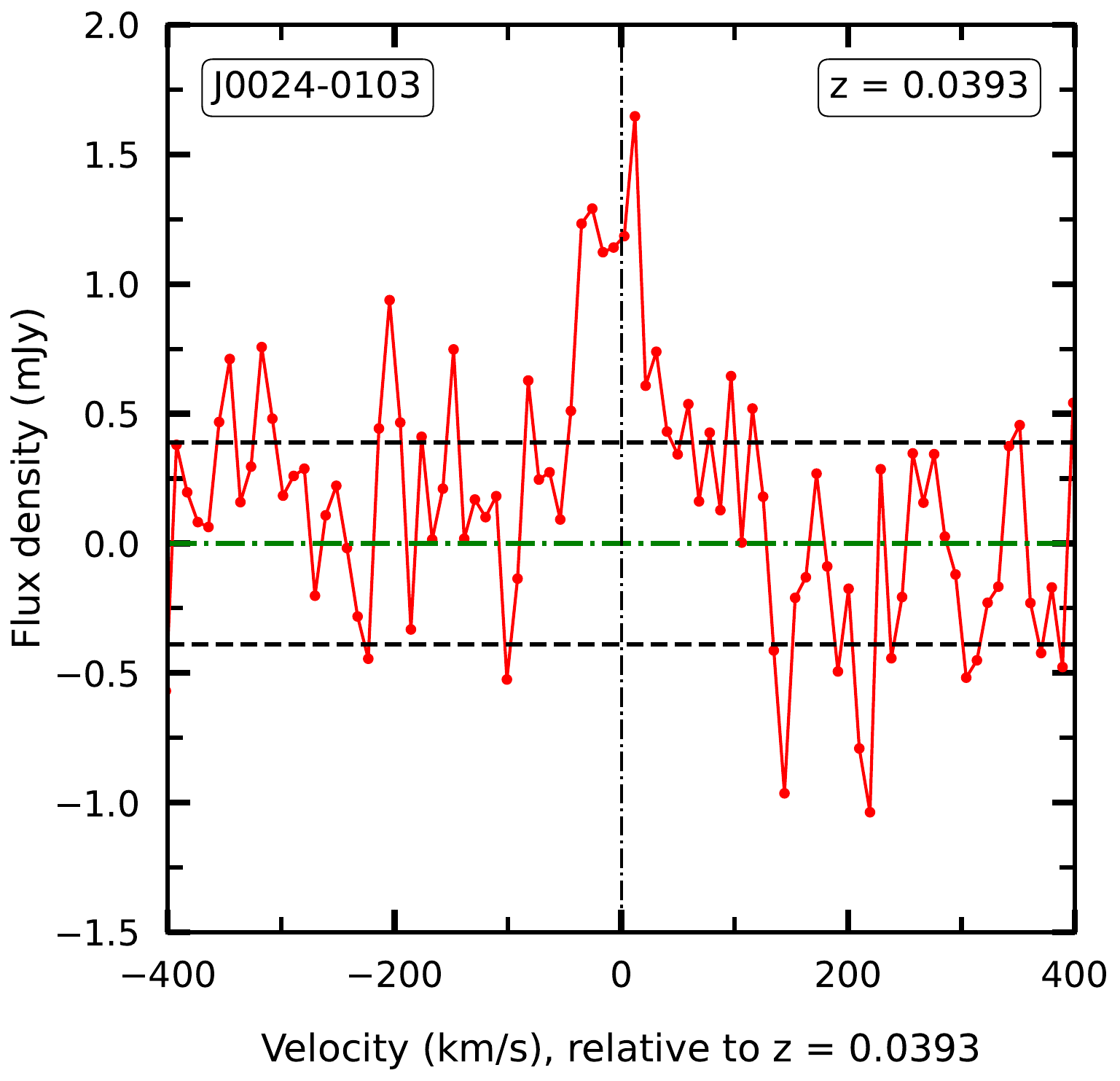}
\includegraphics[width=0.31\textwidth]{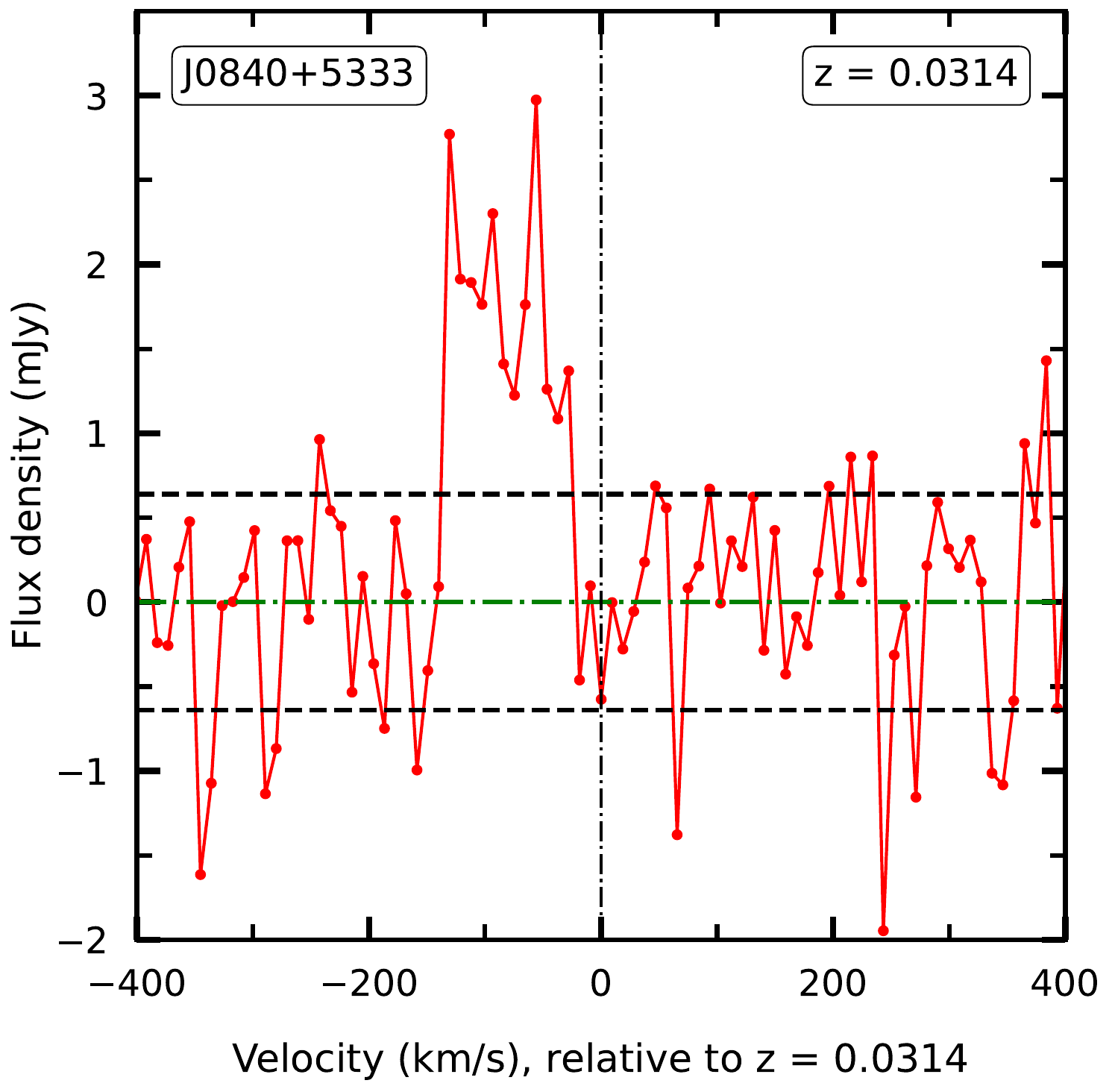}
\includegraphics[width=0.325\textwidth]{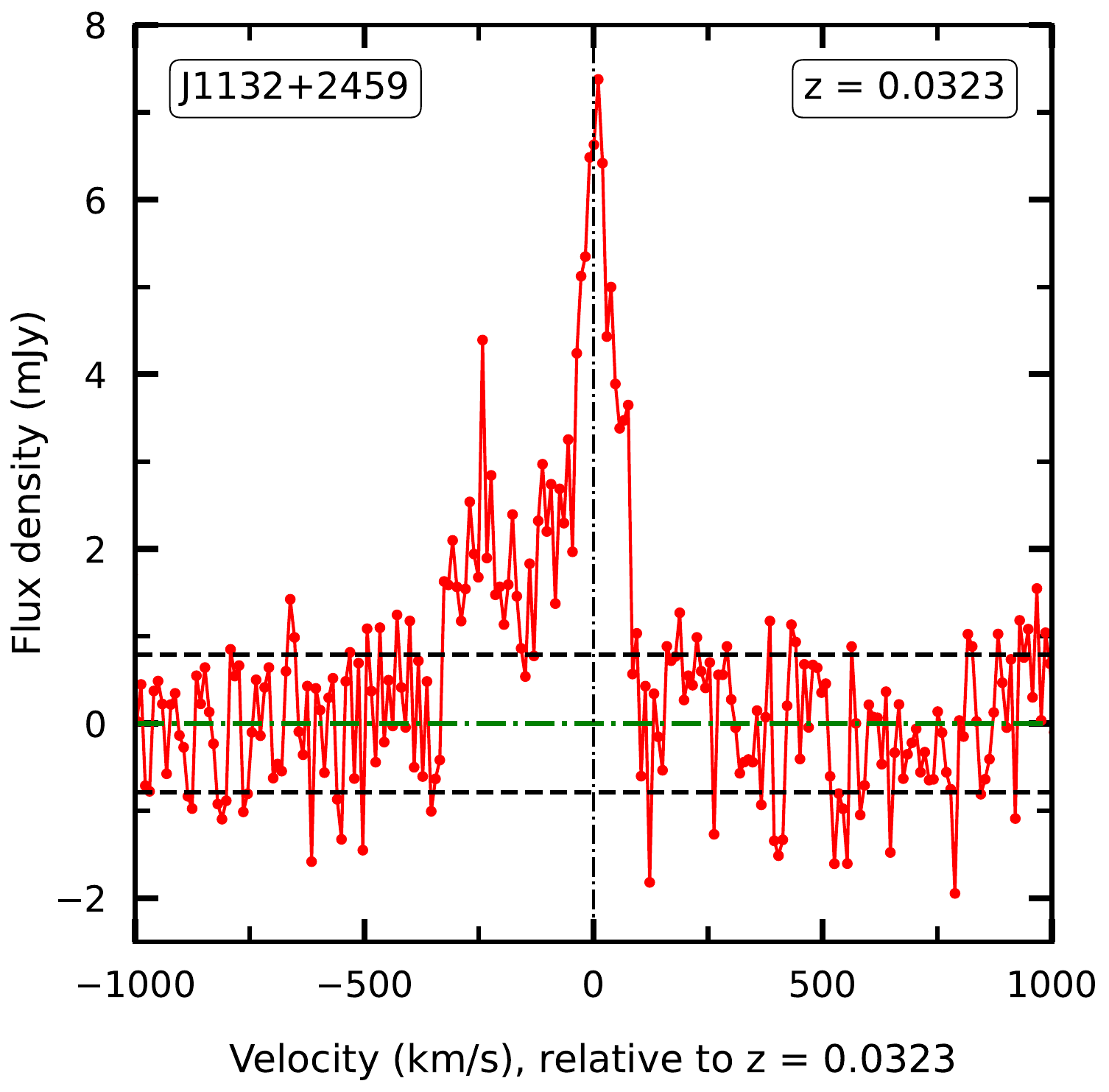}
\includegraphics[width=0.31\textwidth]{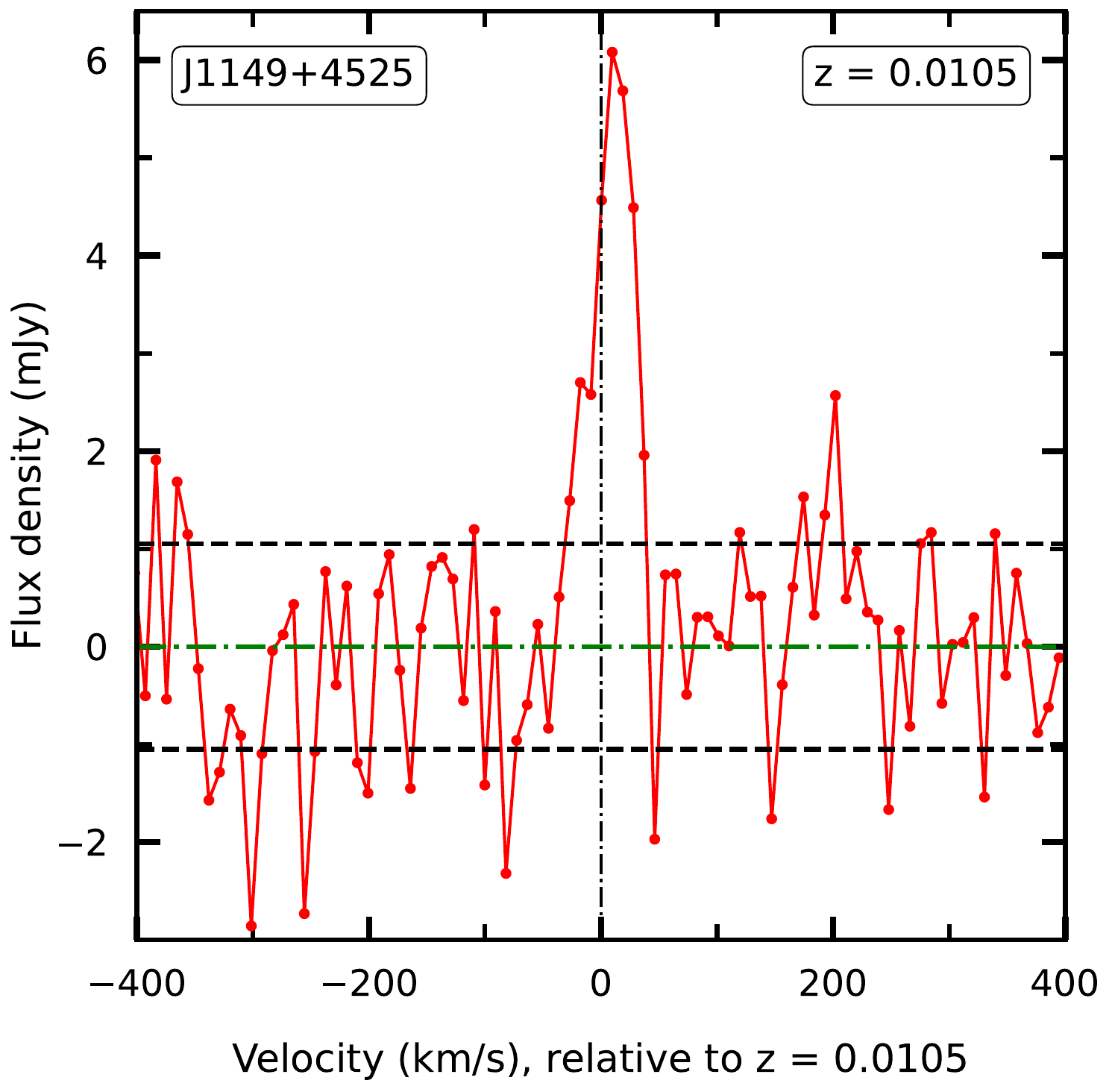}
\includegraphics[width=0.32\textwidth]{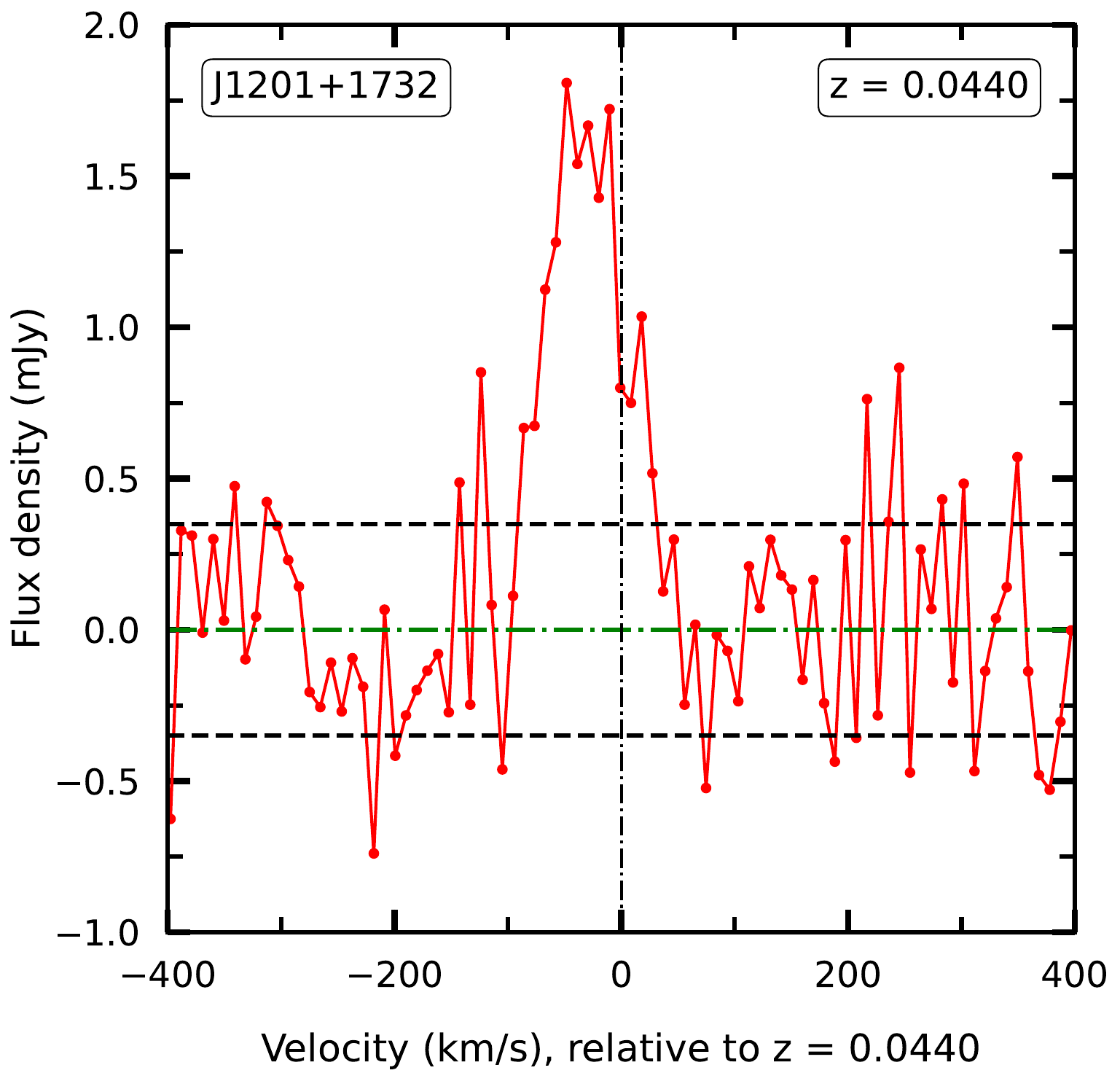}
\includegraphics[width=0.31\textwidth]{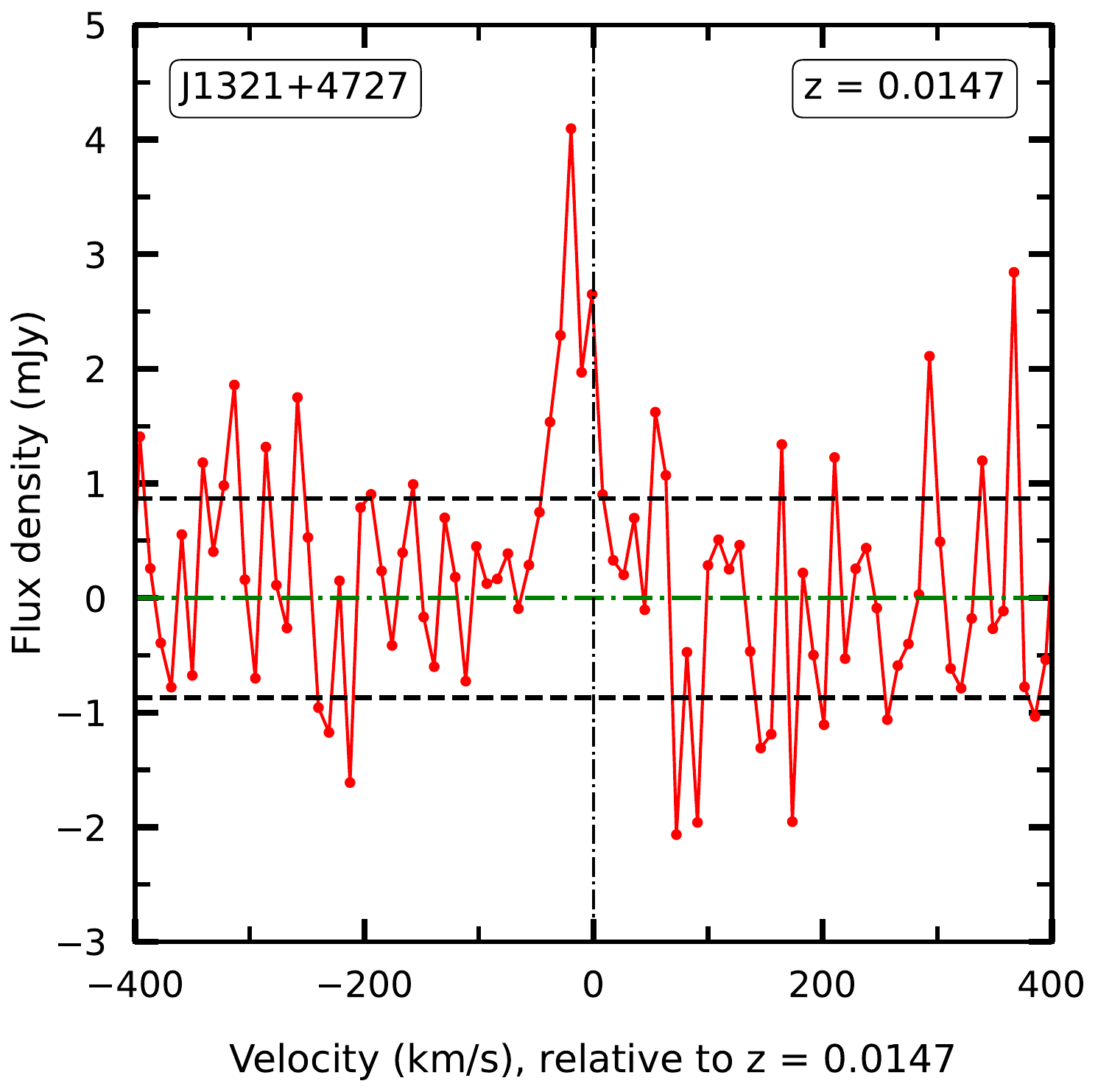}
\includegraphics[width=0.32\textwidth]{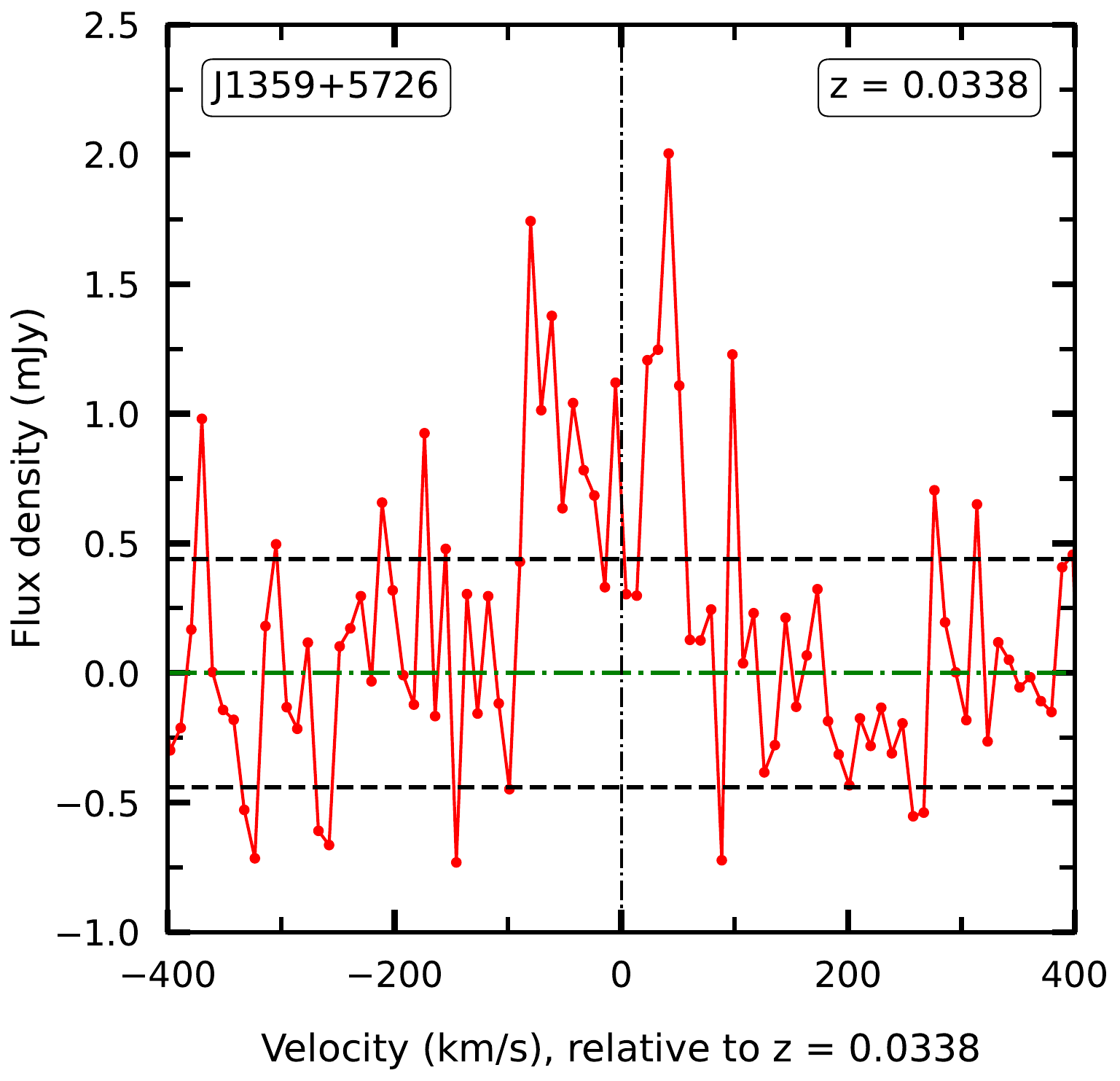}
\caption{GBT \hii\ spectra for the seven GPs of the sample with detections of \hii\ emission, ordered by right ascension. In each panel, the x-axis shows velocity, in \kms, relative to the GP redshift, while the y-axis is flux density, in mJy. The horizontal dashed black lines in each panel indicate the $\pm  1\sigma$ error on the spectrum, at a velocity resolution of $10$~\kms.
\label{fig:detections}}
\end{figure}

\begin{figure}[t!]
\centering
\includegraphics[width=0.24\textwidth]{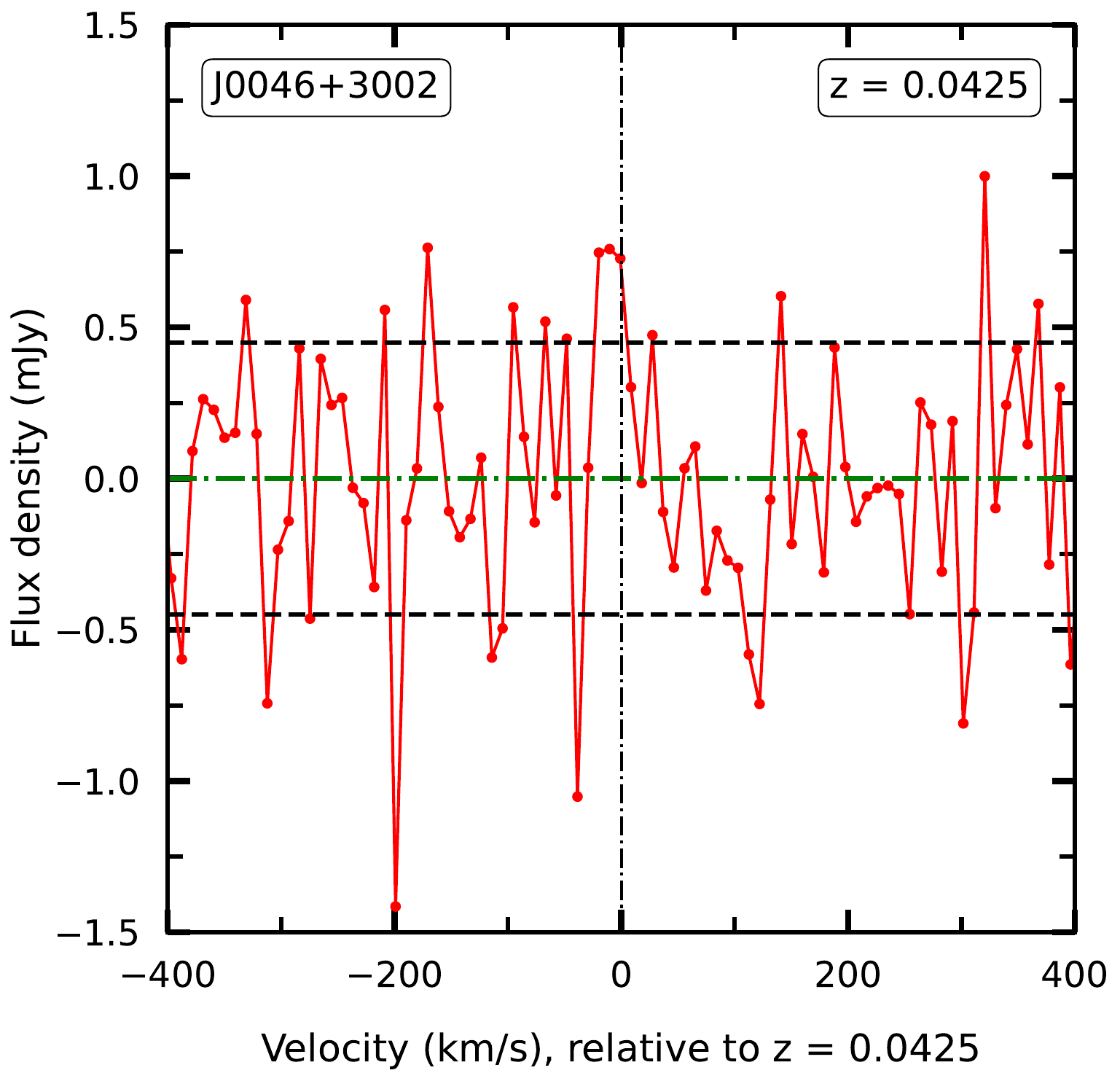}
\includegraphics[width=0.24\textwidth]{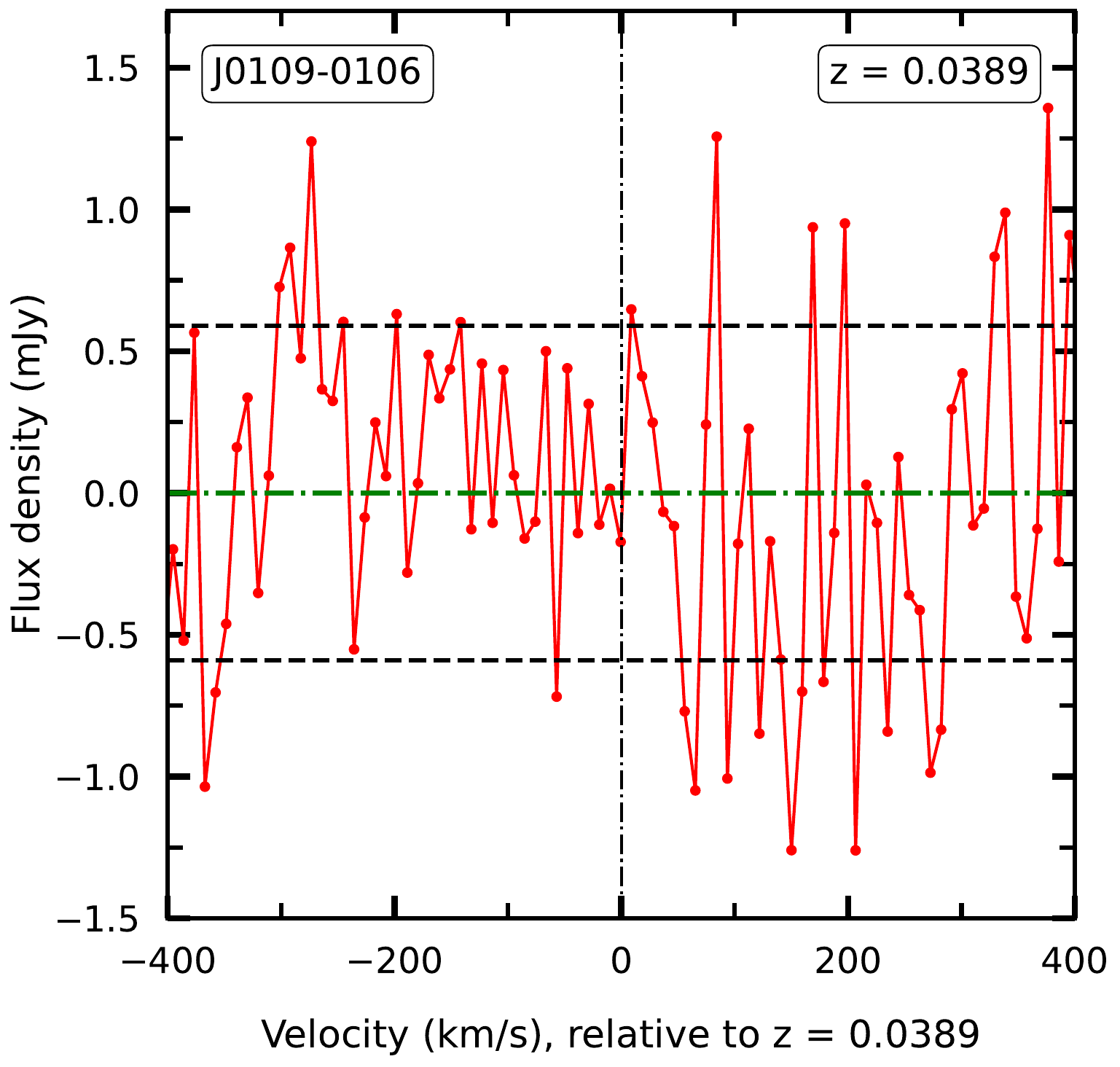}
\includegraphics[width=0.24\textwidth]{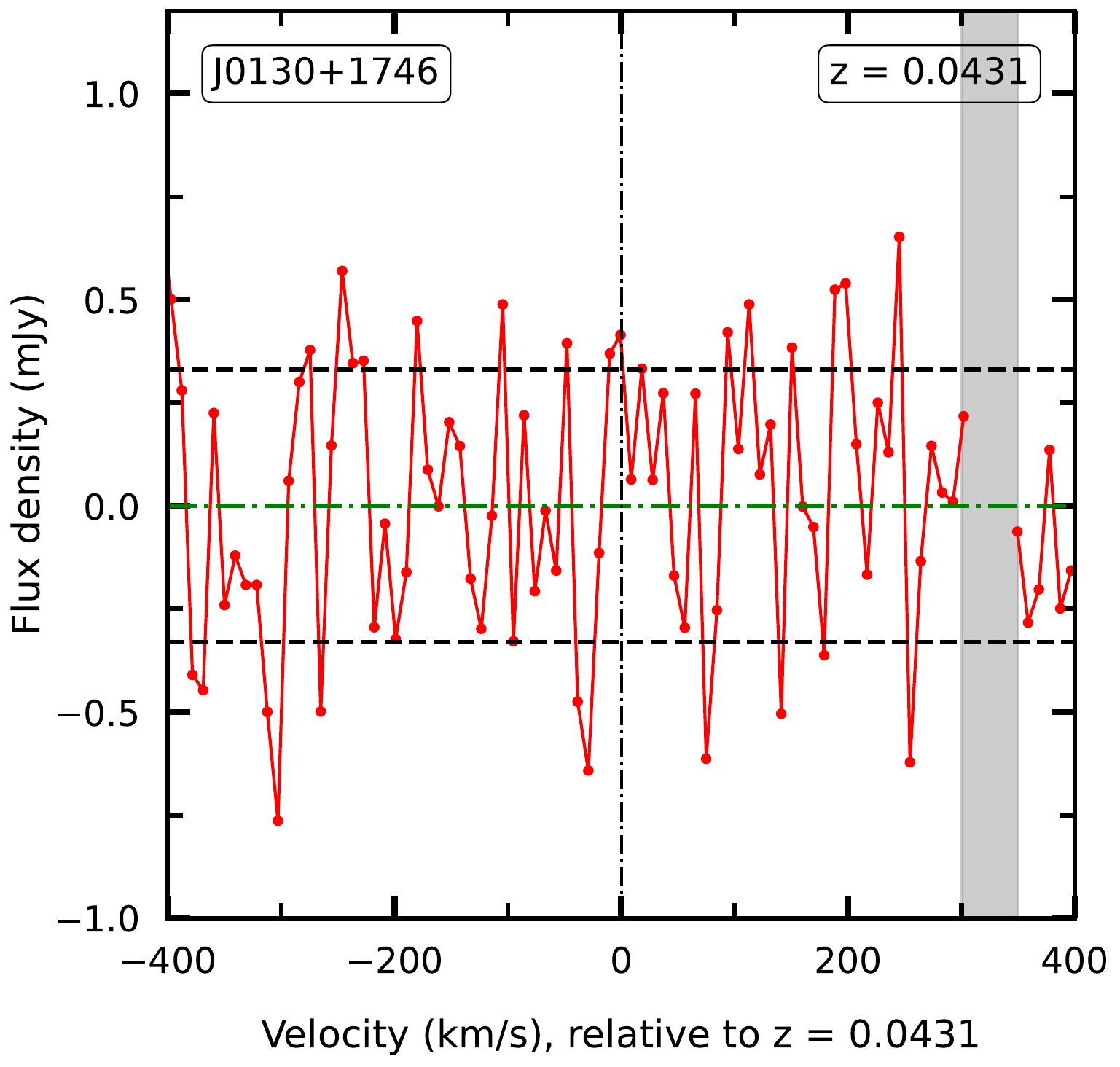}
\includegraphics[width=0.24\textwidth]{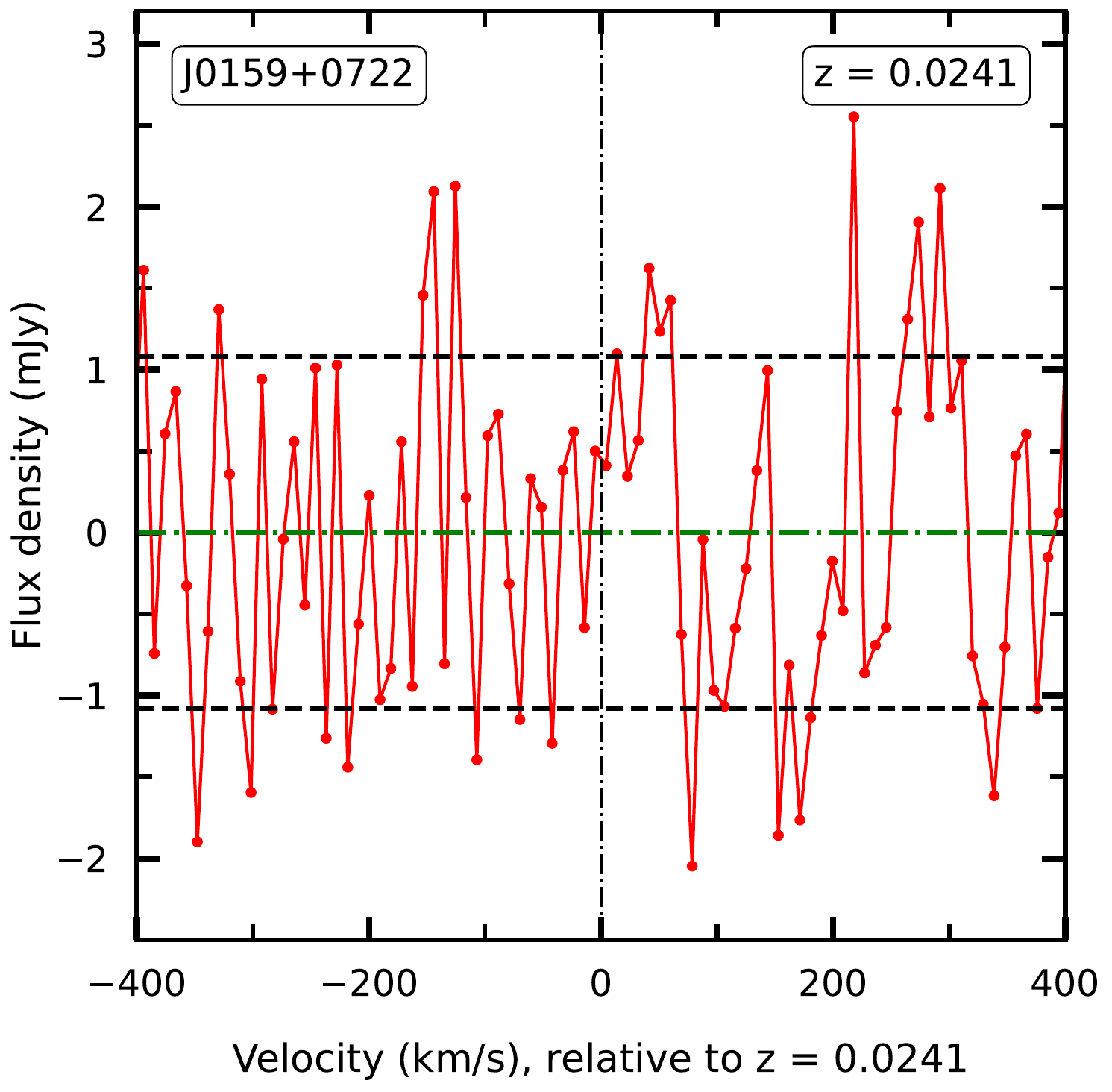}
\includegraphics[width=0.24\textwidth]{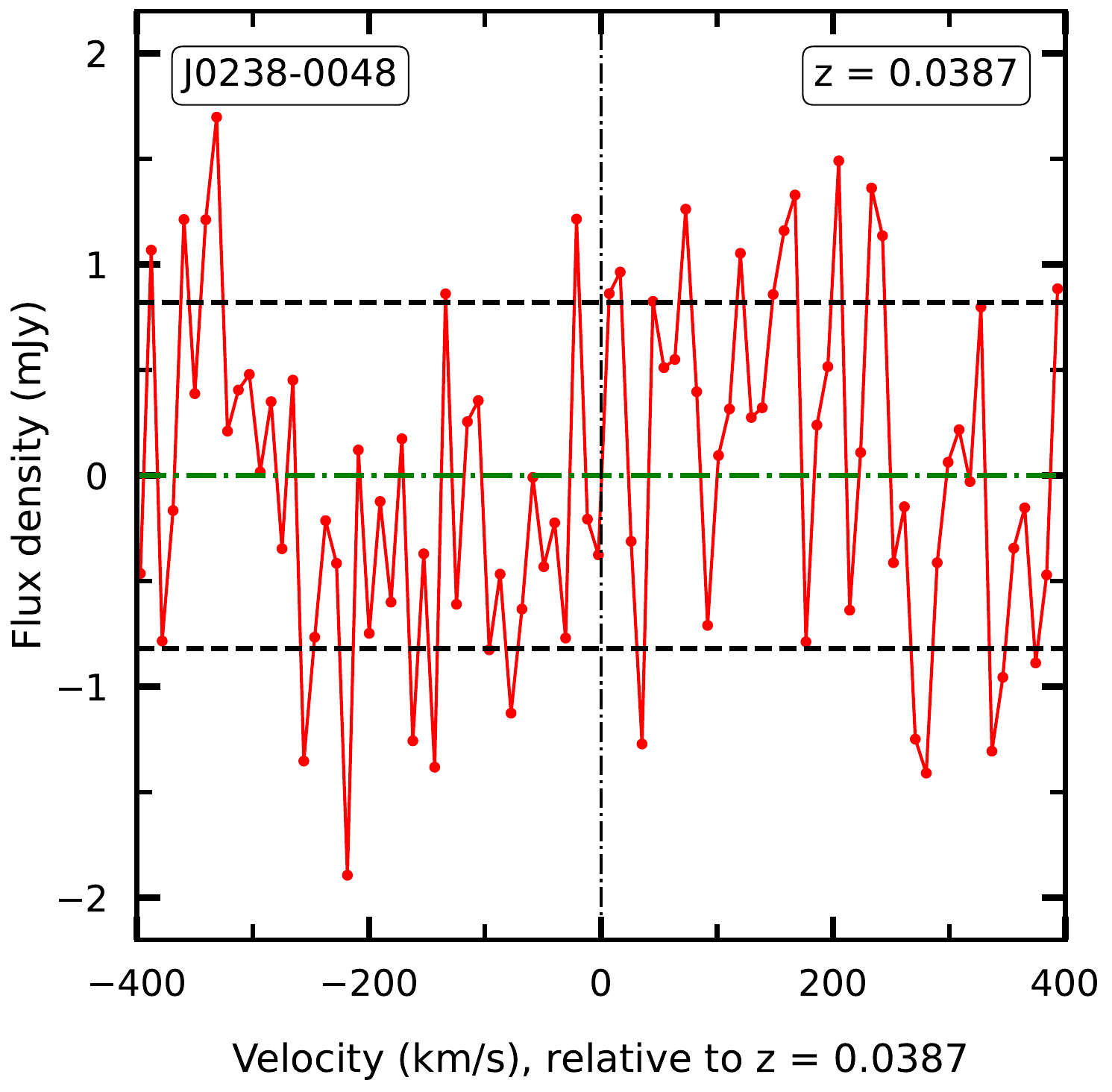}
\includegraphics[width=0.24\textwidth]{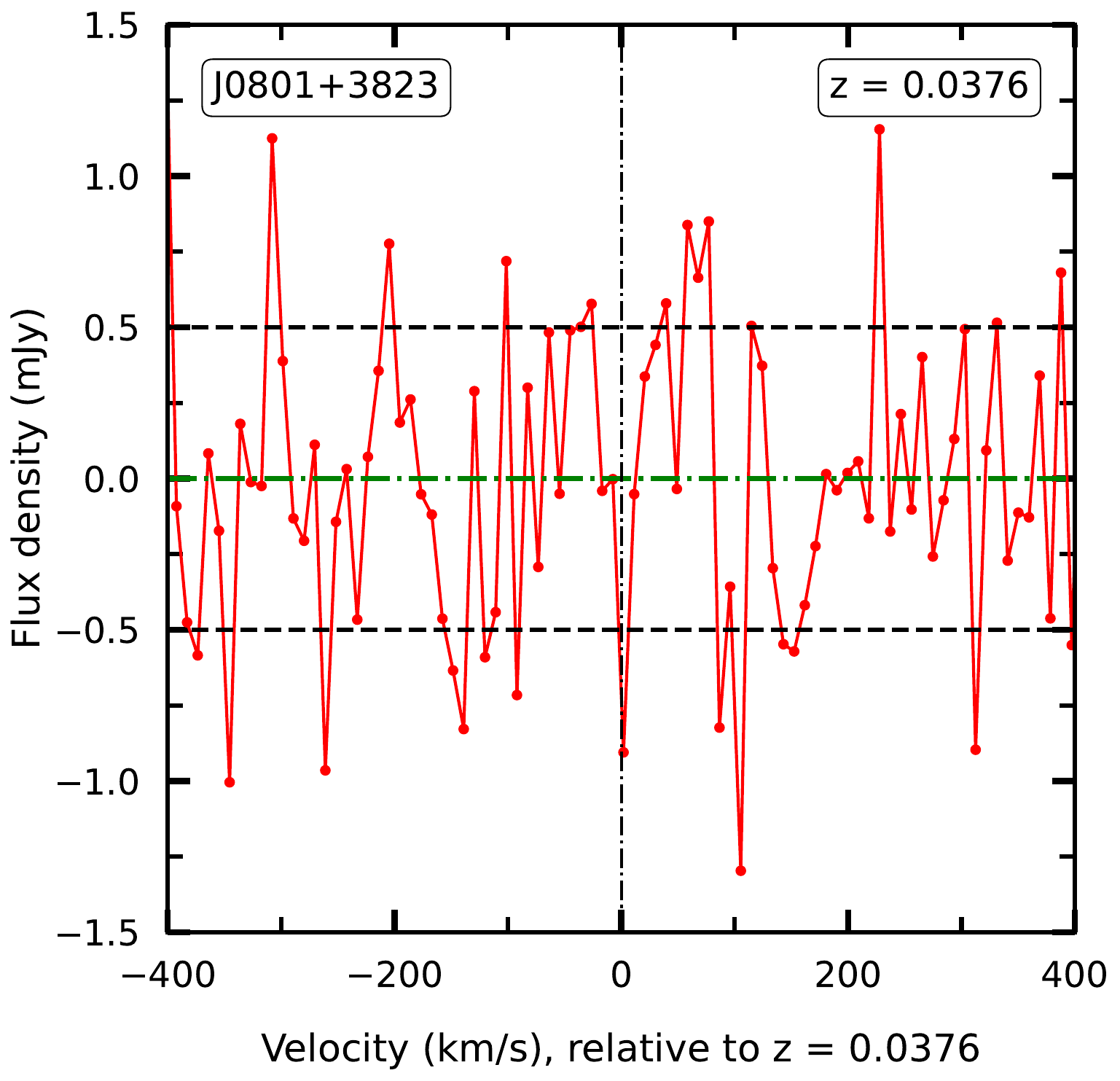}
\includegraphics[width=0.24\textwidth]{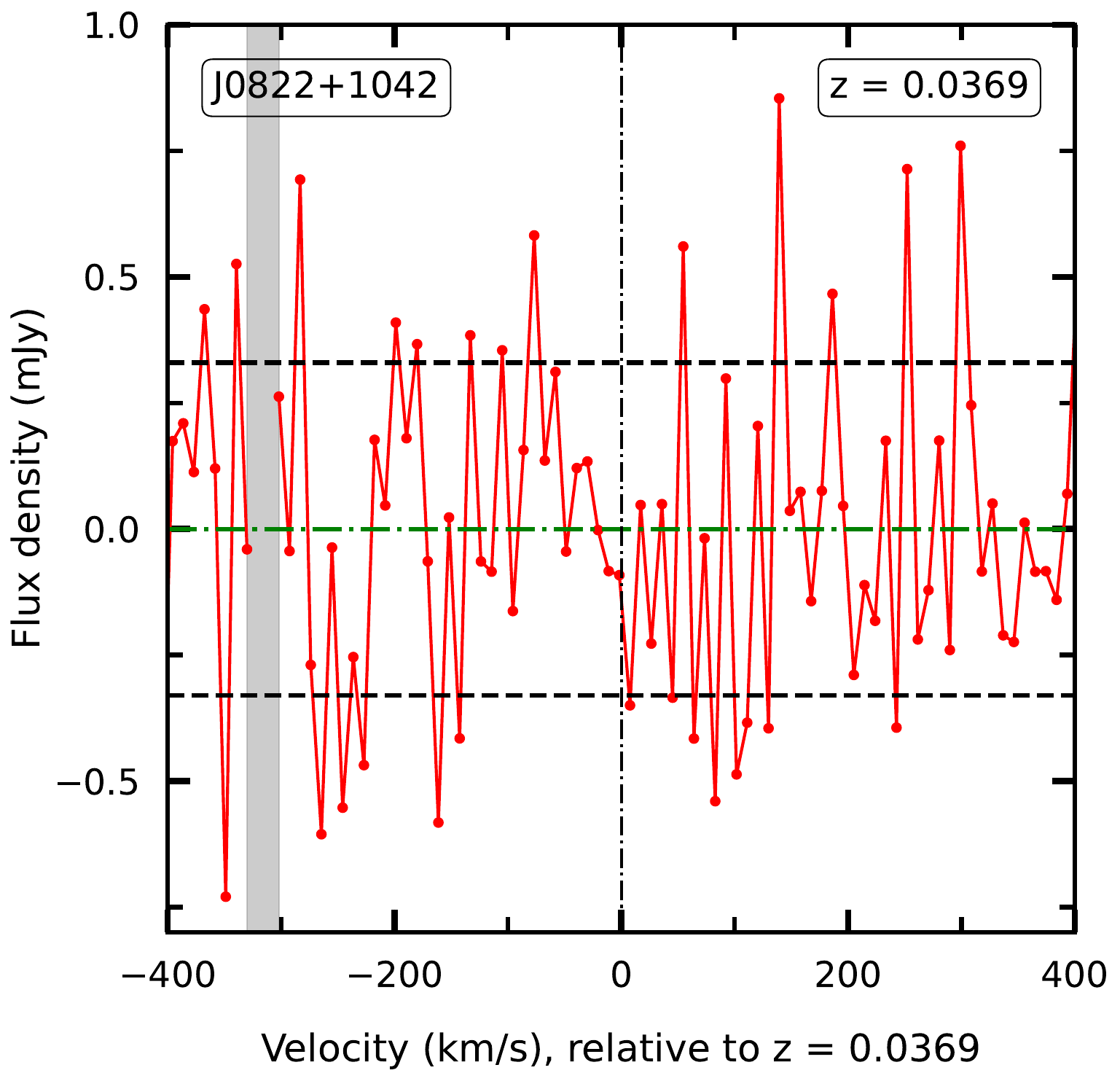}
\includegraphics[width=0.24\textwidth]{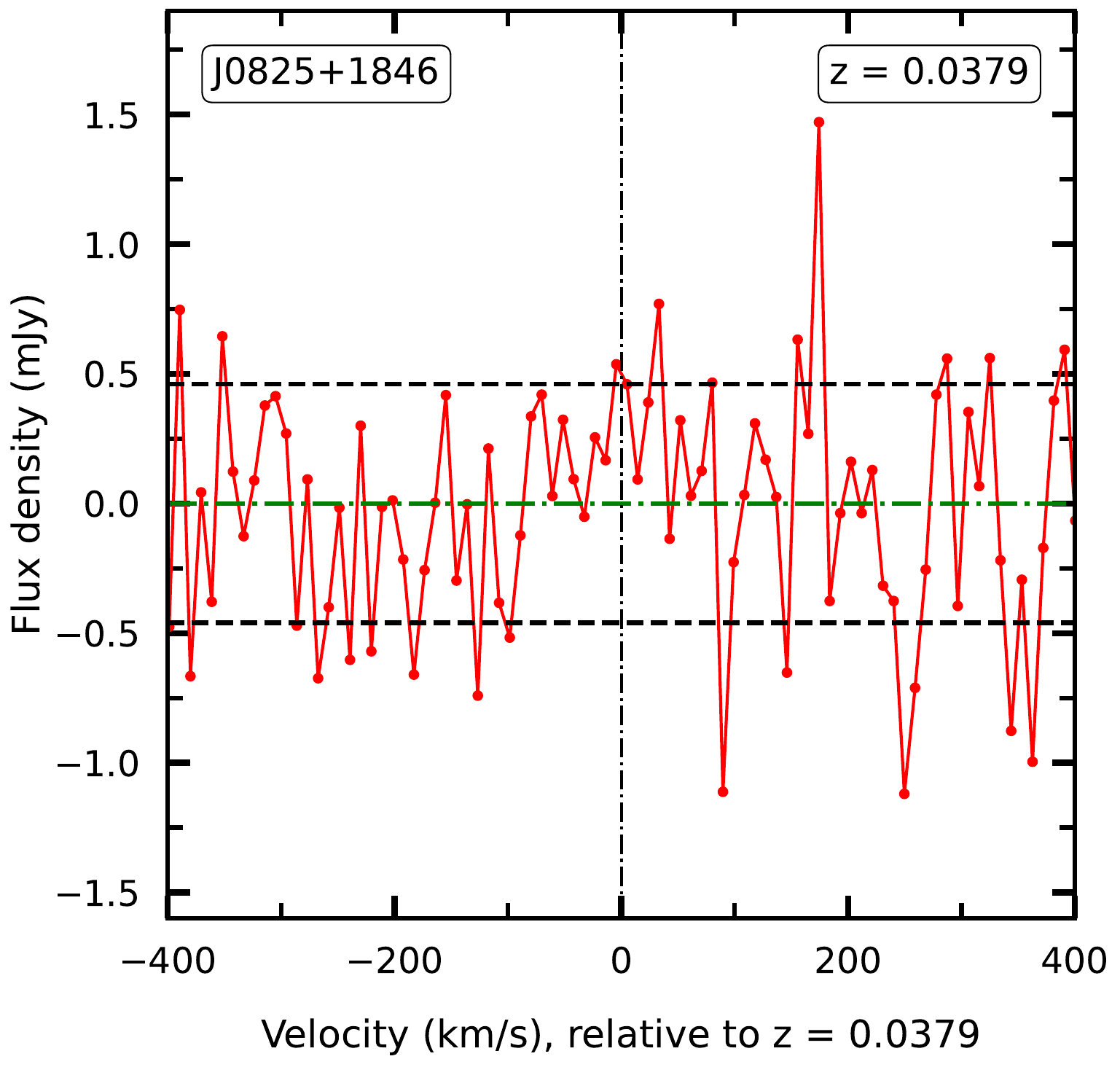}
\includegraphics[width=0.24\textwidth]{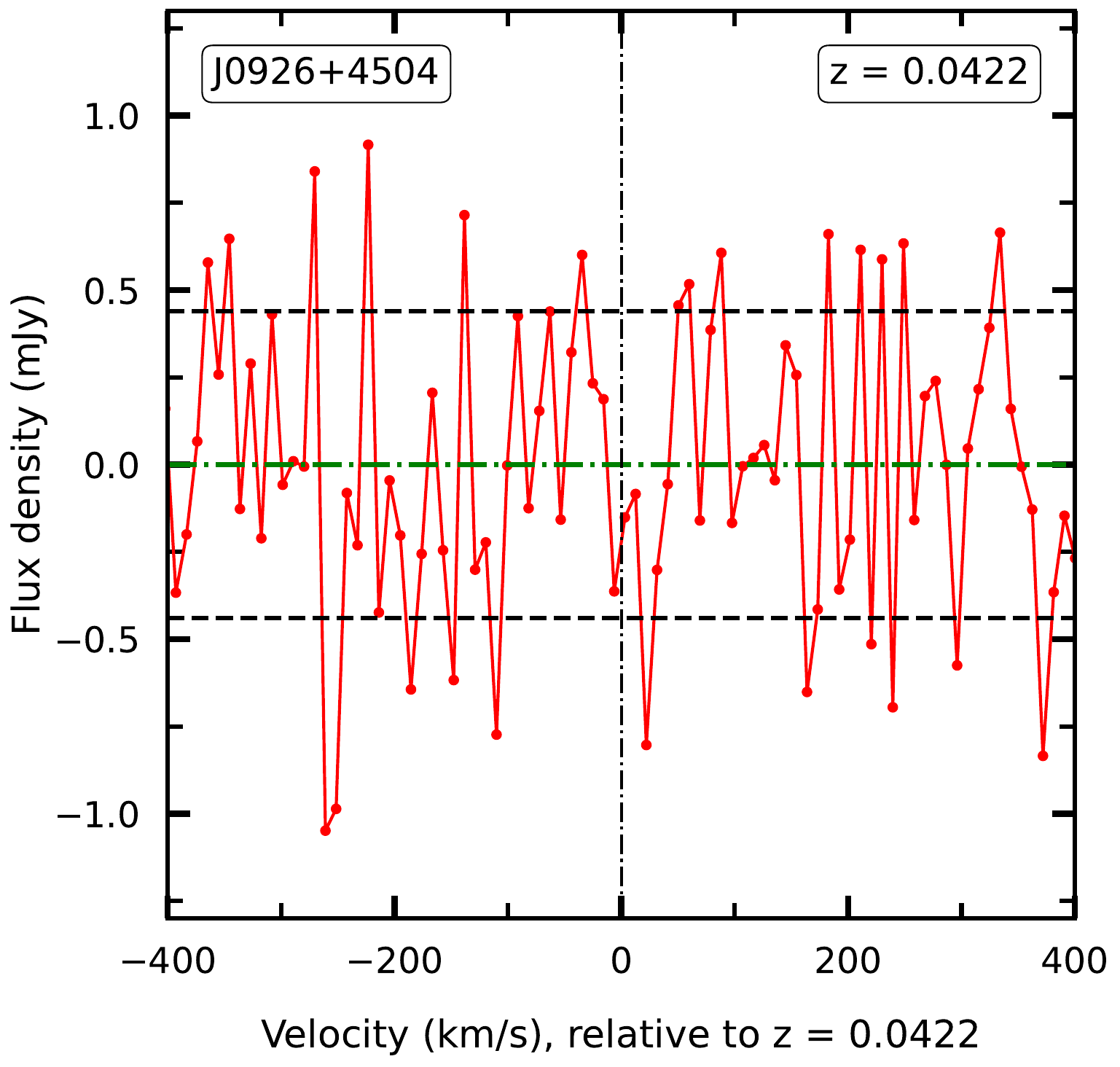}
\includegraphics[width=0.24\textwidth]{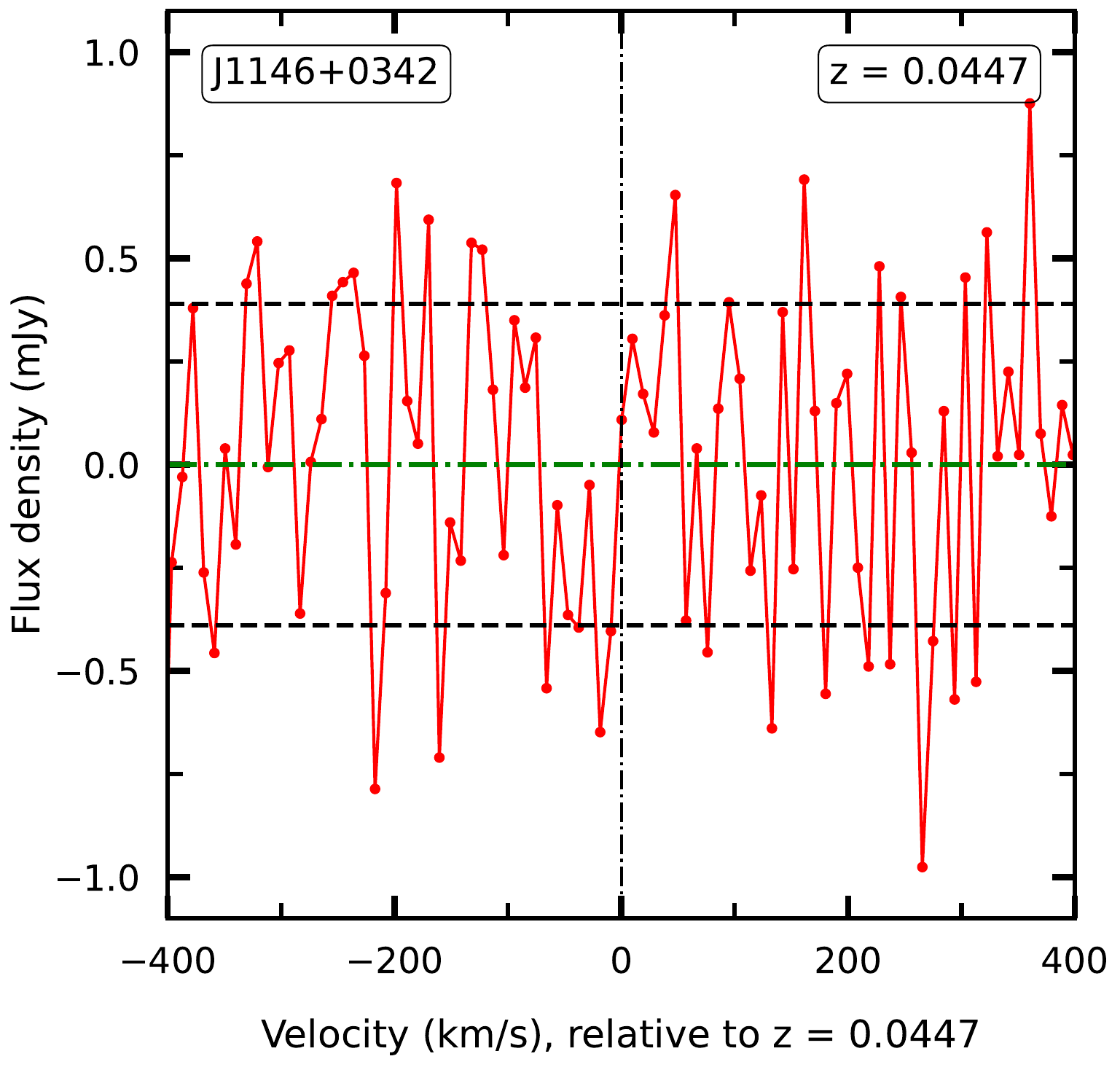}
\includegraphics[width=0.24\textwidth]{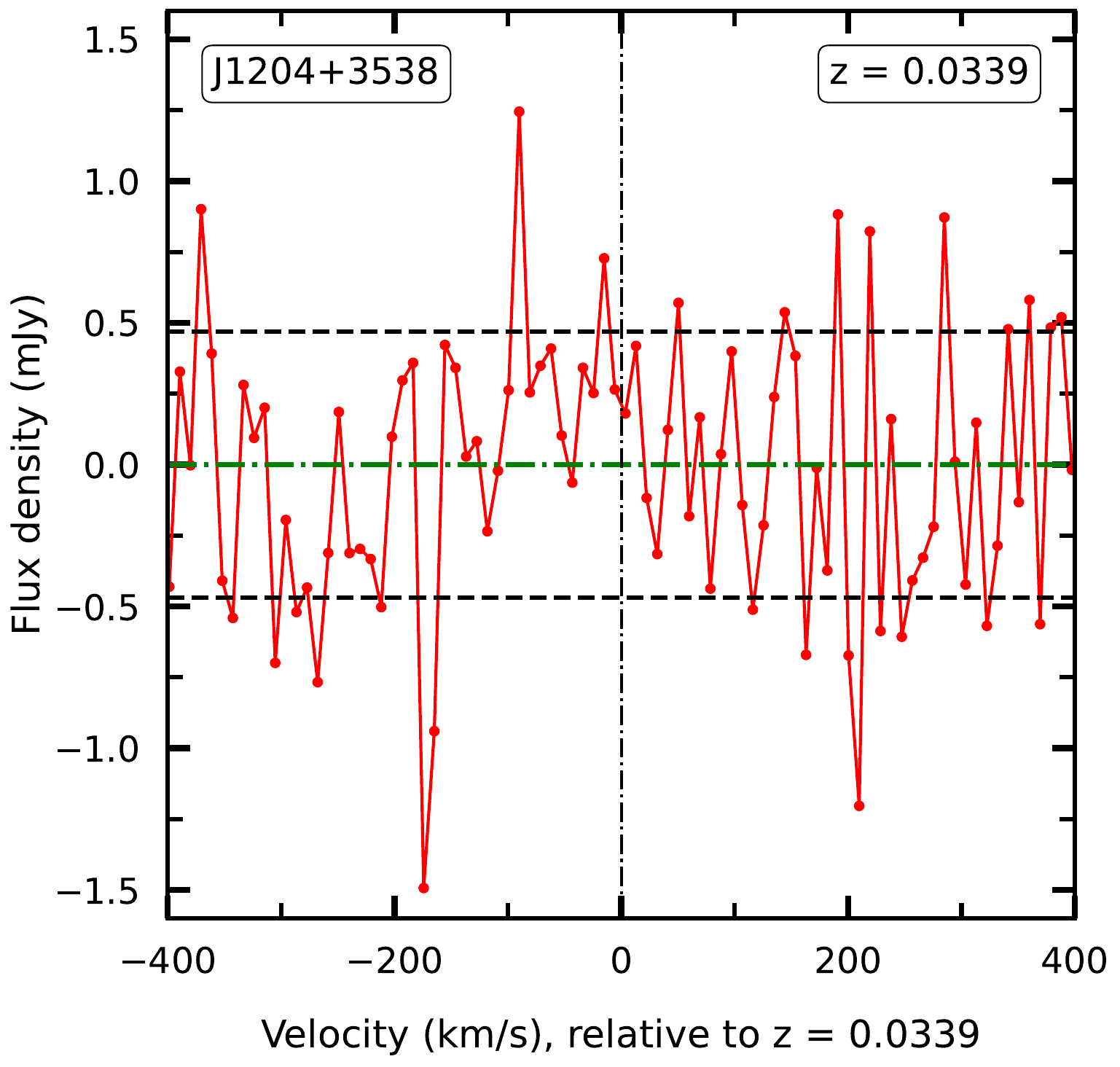}
\includegraphics[width=0.24\textwidth]{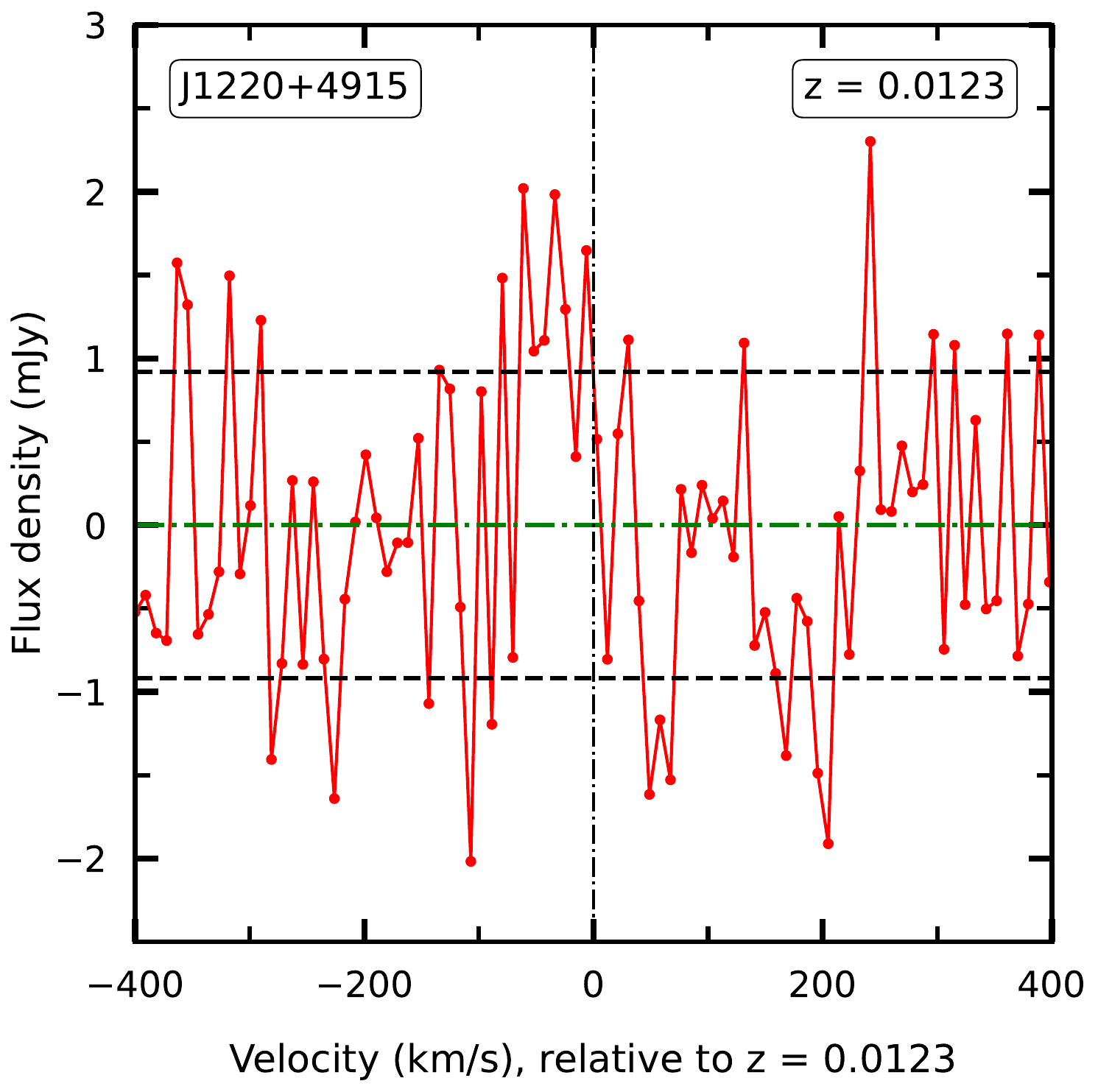}
\includegraphics[width=0.24\textwidth]{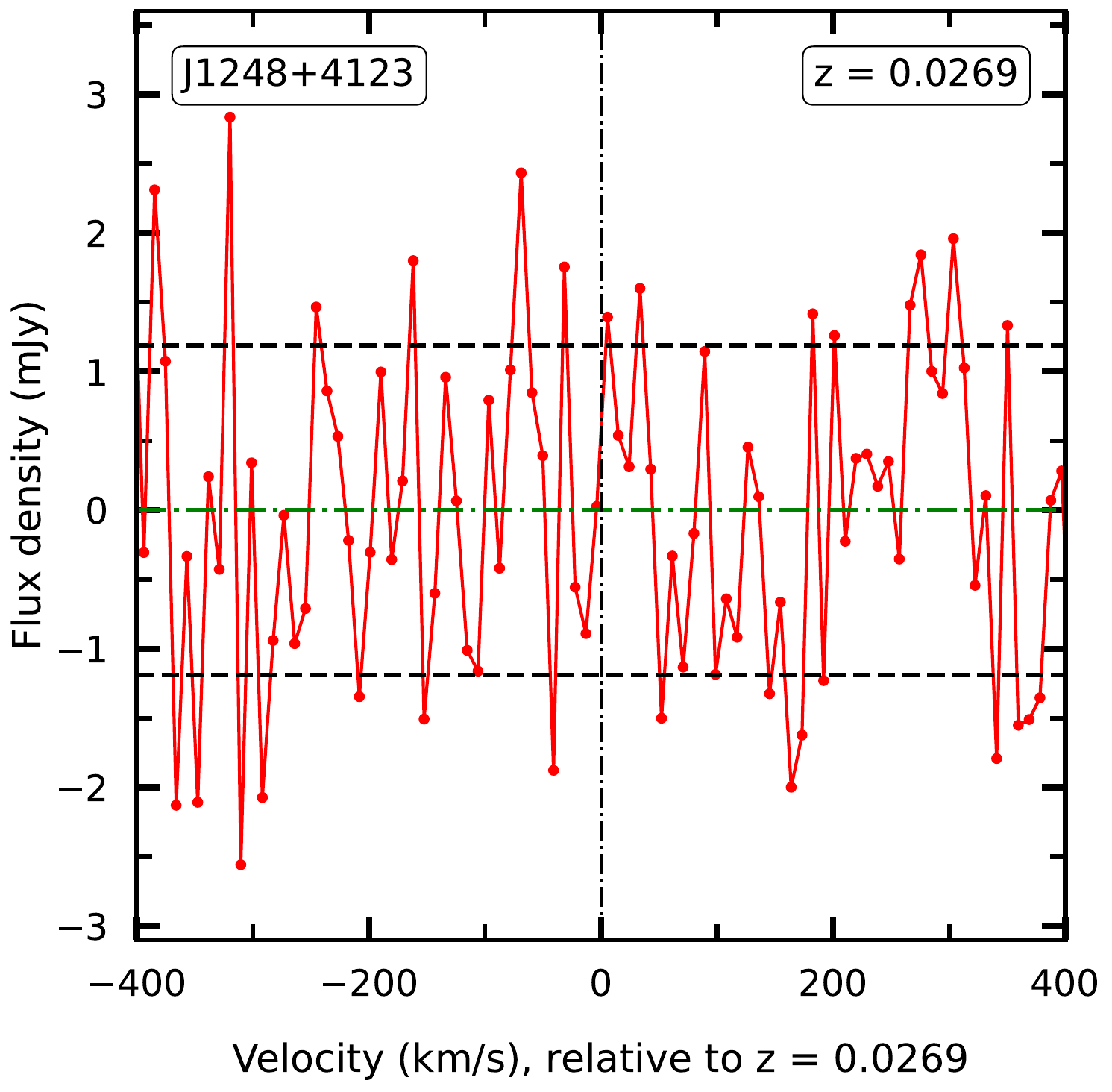}
\includegraphics[width=0.24\textwidth]{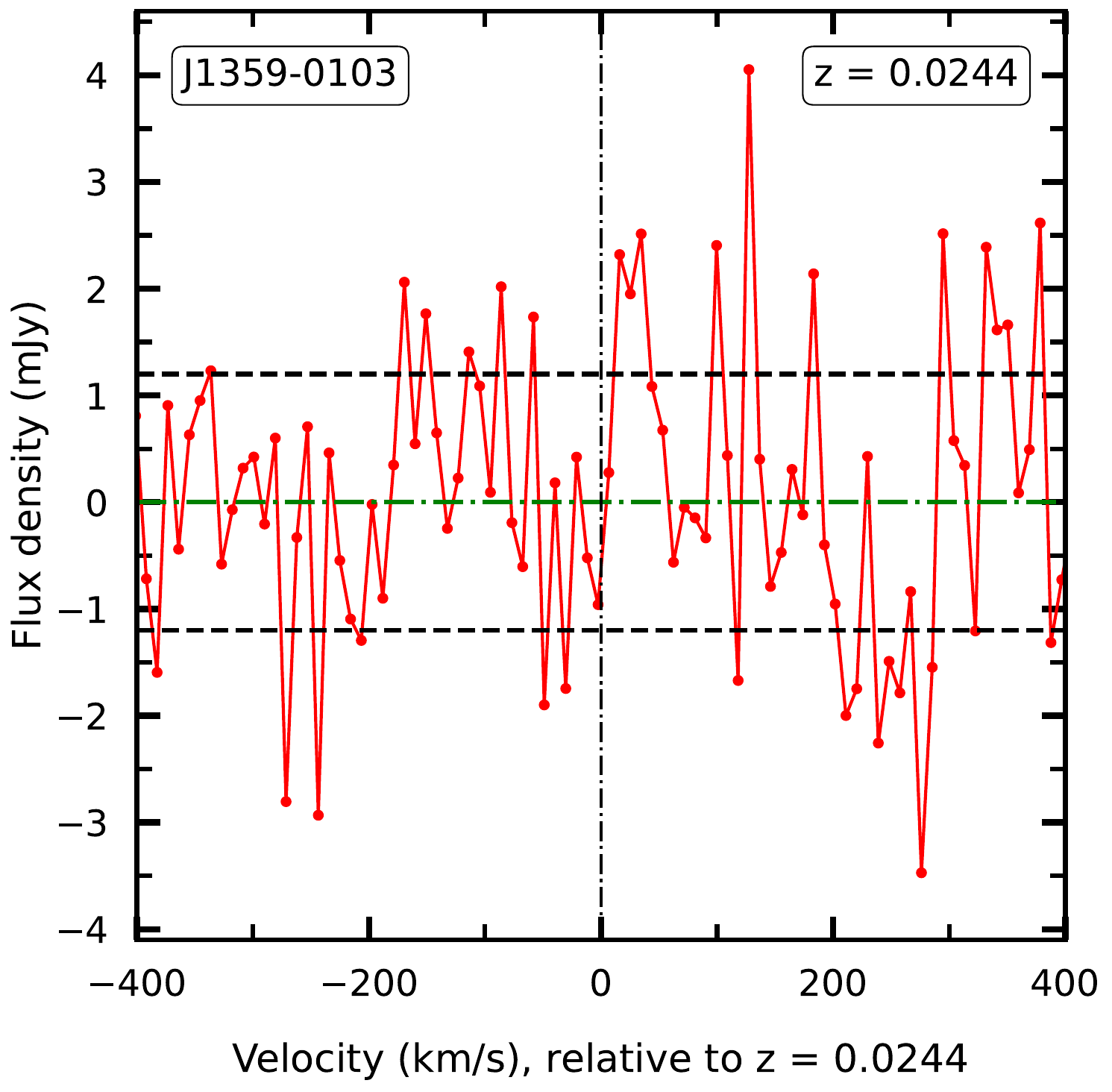}
\includegraphics[width=0.24\textwidth]{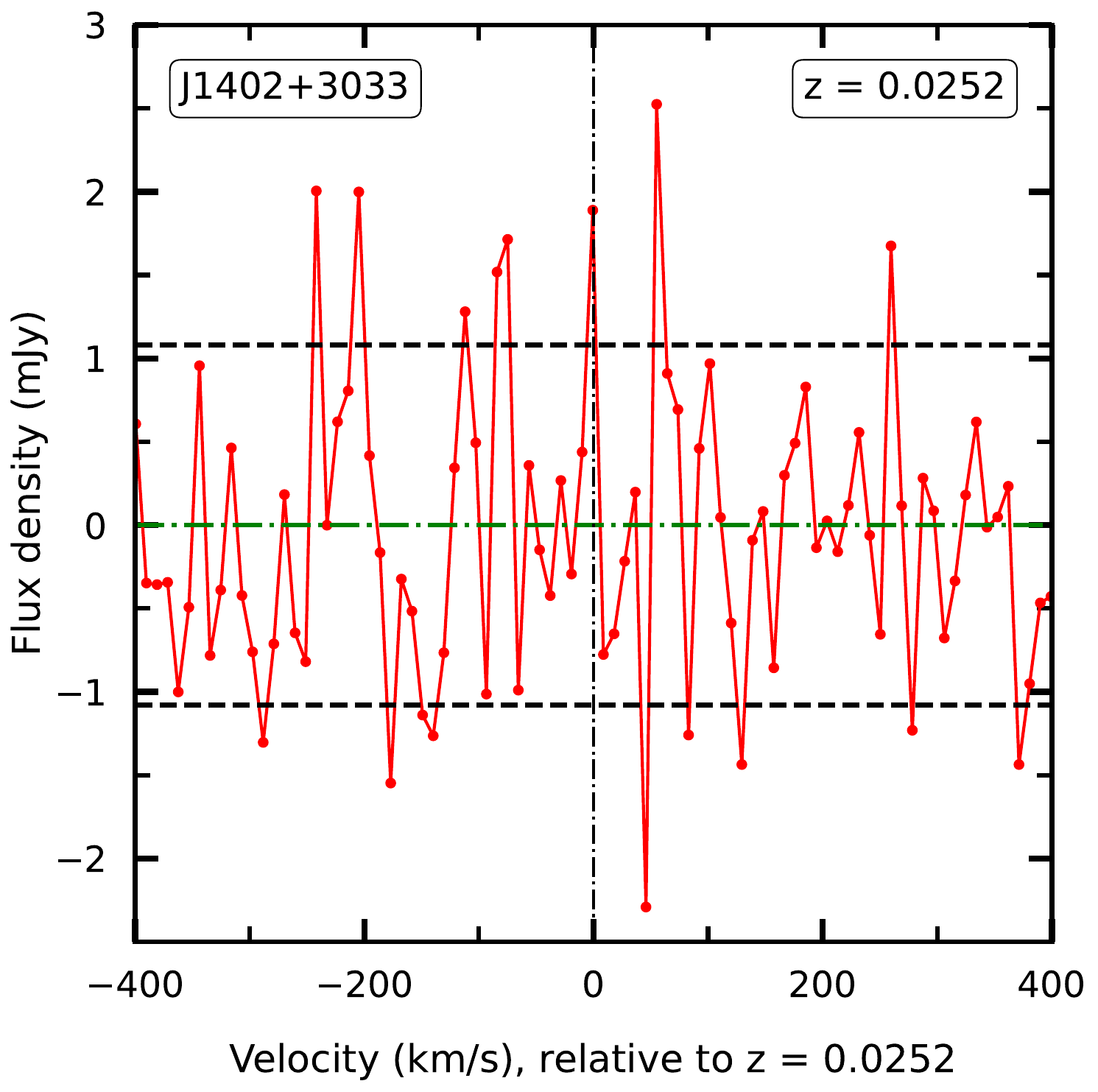}
\includegraphics[width=0.24\textwidth]{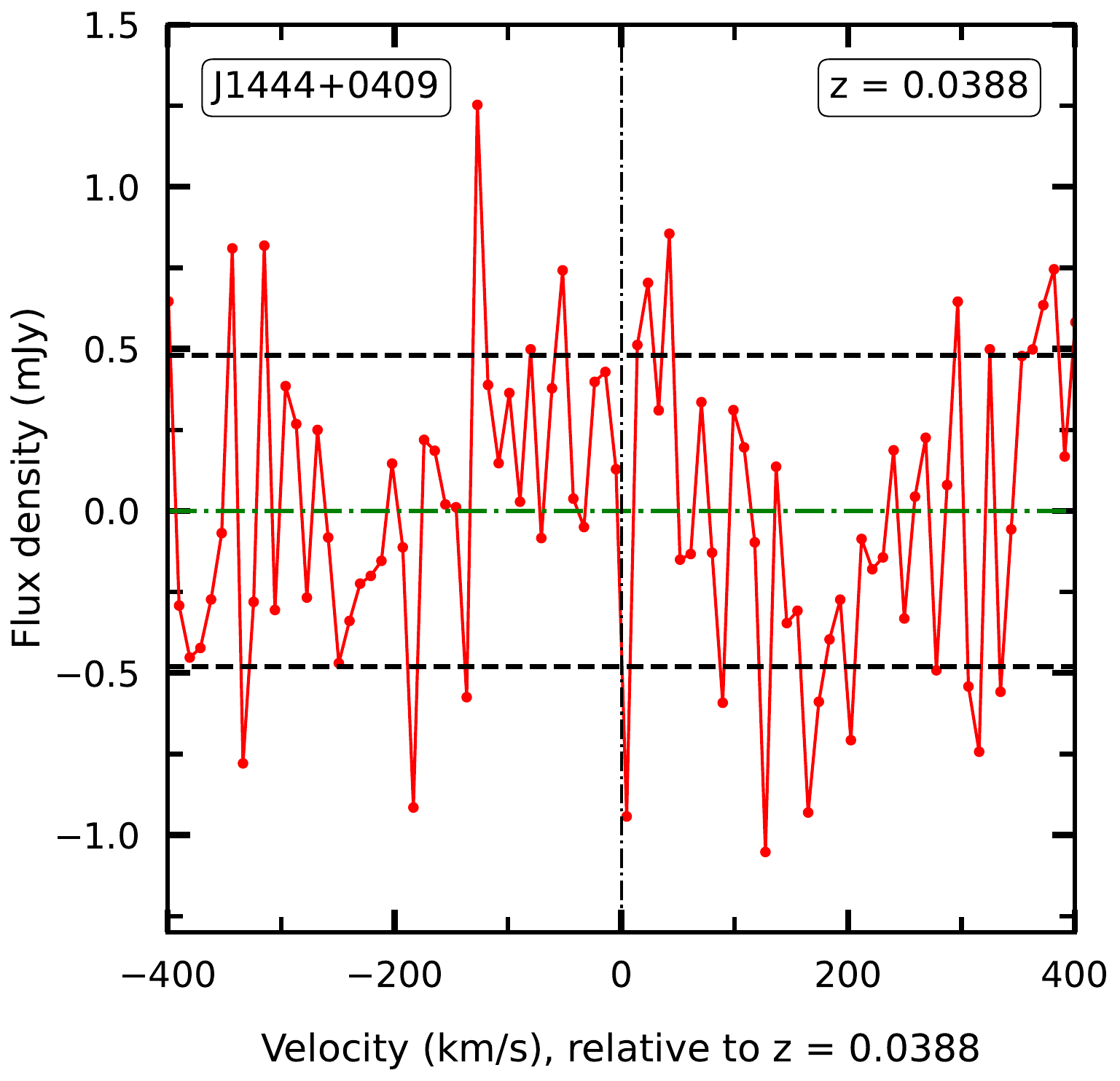}
\includegraphics[width=0.24\textwidth]{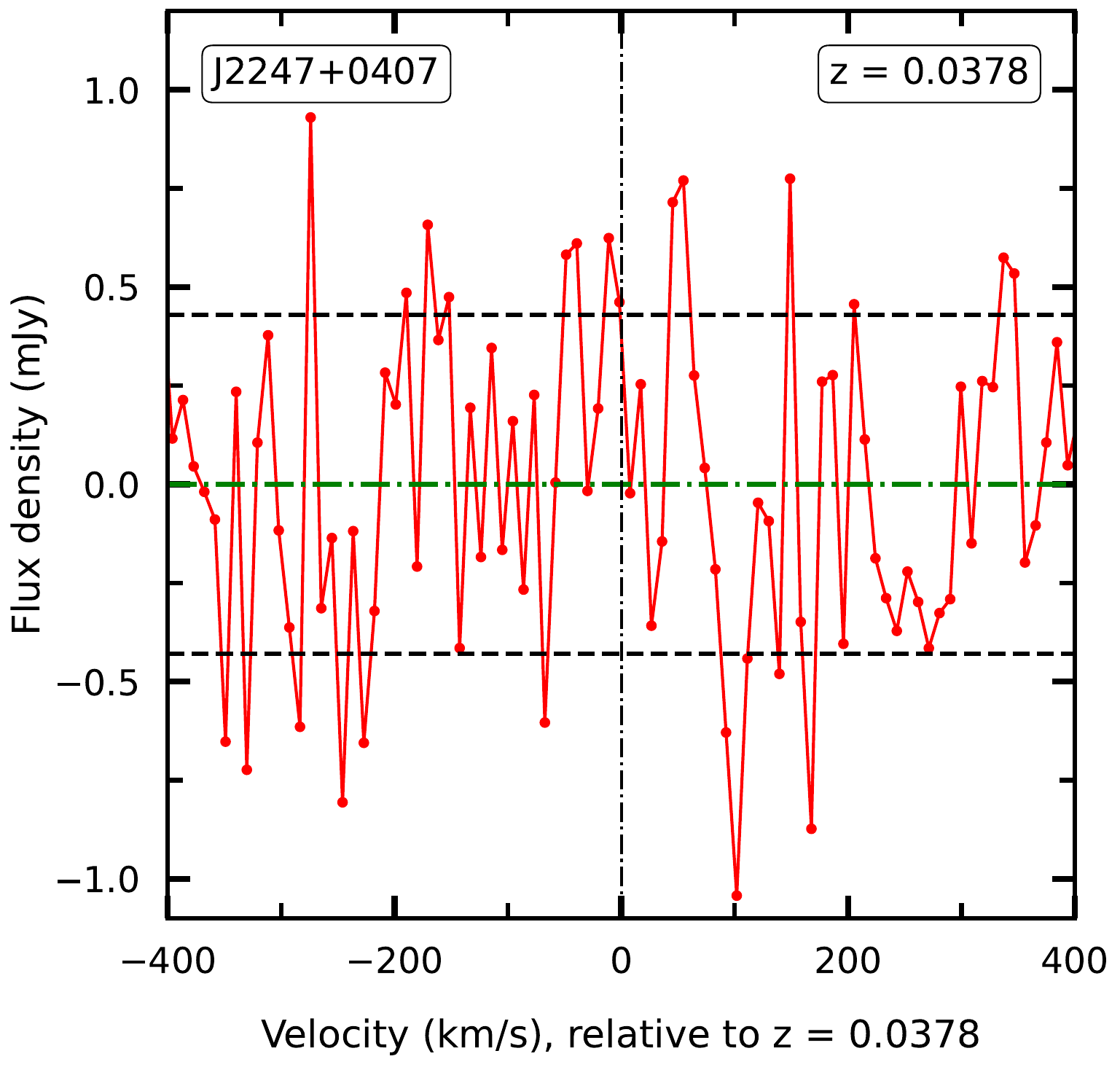}
\caption{GBT \hii\ spectra for the 17 non-detections of the sample, ordered by right ascension. In each panel, the x-axis shows velocity, in \kms, relative to the GP redshift, while the y-axis is flux density, in mJy. The horizontal dashed black lines in each panel indicate the $\pm  1\sigma$ error on the spectrum, at a velocity resolution of $10$~\kms. For J0130+1746 and J0822+1042, the vertical grey bands indicate regions of the spectra that were affected by RFI, and that have not been used in the RMS noise estimates.
\label{fig:non-detections}}
\end{figure}

Our GBT spectra of the 7 \hii\ detections and 17 non-detections are shown in Figures~\ref{fig:detections} and \ref{fig:non-detections}, respectively. For most of the 7 detections, the \hii\ emission lies close to the expected redshift, based on the optical lines. However, for J0840+5333, the \hii\ emission is clearly offset from the optical redshift, by $\approx -100$~\kms; it is possible that this \hii\ signal arises from a companion galaxy of the GP, also lying within the $\approx 9'$ GBT beam.

For each \hii\ spectrum, we estimated the H{\sc i} mass of the GP via the relation 
\begin{equation}
{\rm M}_{\text{HI}} =2.36 \times 10^5 \times {\rm D_L^2} \times \frac{\rm \int SdV}{(1+z)} \,\, ,
\label{eqn:hi}
\end{equation}
where $\rm M_{\text{HI}}$ is in $\rm M_{\odot}$, the luminosity distance, $\rm D_L$, is in Mpc and the velocity-integrated \hii\ line flux density, $\rm \int SdV$, is in Jy~\kms. For the 17 non-detections, we assumed a Gaussian line profile with a line FWHM of 50~\kms, typical of dwarf galaxies, smoothed each \hii\ spectrum to a velocity resolution of 50~\kms, and measured the RMS noise at this resolution. The resulting $3\sigma$ upper limits on the velocity-integrated line flux densities were then used to obtain upper limits to the H{\sc i} masses of the GPs, using Equation~\ref{eqn:hi}.

For the GPs of our sample, Table~\ref{tab:gp-data} lists, in order of right ascension, the GP name, the redshift, the expected redshifted \hii\ line frequency (in MHz), the velocity-integrated \hii\ line flux density and error, or 3$\sigma$ upper limits to the \hii\ line flux density (in Jy~\kms), the inferred H{\sc i} mass and error, or 3$\sigma$ upper limits to the \hi\ mass (in $\rm 10^8 \, \Msun$), the star formation rate \citep[SFR, in $\rm \Msun$~yr$^{-1}$; ][]{jiang19}, the stellar mass $\rm M_*$ \citep[in $\rm 10^8 \, \Msun$; ][]{jiang19}, the H{\sc i}-to-stellar mass ratio, $\rm f_{HI} = M_{HI}/M_{*}$, the H{\sc i} depletion timescale, $\rm \tau_{dep} = M_{HI}/SFR$ (in Gyr), the absolute B-band magnitudes, $\rm M_B$, the O32 ratio, and the FWHM of the \hii\ line for the \hii\ detections ($\rm W_{50}$, in \kms).

\begin{table}[t!]
\centering
\begin{tabular}{|l|c|c|c|c|c|c|c|c|c|c|c|}
\hline
\hline
    Green Pea  &  $z$ & $\rm \nu_{21cm}$ & $\rm \int SdV$ & $\rm M_{HI}$ & SFR & $\rm M_{*}$ & $\rm f_{HI}$ & $\rm \tau_{dep} $ & $\rm M_B$ & O32 & $\rm W_{50}$ \\
    identifier & \, & (MHz) & (Jy km/s) & ($\rm \times 10^8 \, M_{\odot}$) & ($\rm M_{\odot}$/yr) & ($\rm \times 10^8 \, M_{\odot}$) & \, & (Gyr) & \, & \, & (km/s) \\
    \hline
     J0024-0103 & 0.0393 & 1366.63  & $0.127 \pm 0.016$ & $9.3 \pm 1.2$ & 0.37 & 0.25 & 34 & 2.6 & -16.89 & 8.0 & 85 \\
     \hline
     J0046+3002 & 0.0425 & 1362.47  & $< 0.034$ & $< 3.0$ & 0.048 & 0.01 & $< 245$ & $< 6.2$ & -15.43 & 3.7 & - \\
     \hline
     J0109-0106 & 0.0389 & 1367.20  & $< 0.056$ & $< 4.2$ & 0.17 & 4.0 & $< 1.1$ & $< 2.4$ & -15.98 & 10.1 & - \\
     \hline
     J0130+1746 & 0.0431 & 1361.77  & $< 0.030$ & $< 2.7$ & 0.049 & 1.6 & $< 1.7$ & $< 5.4$ & -14.90 & 5.1 & - \\
     \hline
     J0159+0722 & 0.0241 & 1386.98  & $< 0.121$ & $< 3.4$ & 0.005 & 0.0016 & $< 2154$ & $< 68$ & -12.76 & 6.4 & - \\
     \hline
     J0238-0048 & 0.0387 & 1367.53  & $< 0.078$ & $< 5.5$ & 0.059 & 2.0 & $< 2.9$ & $< 9.4$ & -15.88 & 12.5 & - \\
     \hline
     J0801+3823 & 0.0376 & 1368.89  & $< 0.046$ & $< 3.1$ & 0.57 & 15.9 & $< 0.2$ & $< 0.5$ & -16.08 & 3.4 & - \\
     \hline
     J0822+1042 & 0.0369 & 1369.90  & $< 0.026$ & $< 1.7$ & 0.046 & 0.63 & $< 2.8$ & $< 3.6$ & -14.56 & 7.3 & - \\
     \hline
     J0825+1846 & 0.0379 & 1368.52  & $< 0.034$ & $< 2.3$ & 0.68 & 0.016 & $< 139$ & $< 0.3$ & -16.90 & 15.8 & - \\
     \hline
     J0840+5333 & 0.0314 & 1377.19  & $0.204 \pm 0.020$ & $9.5 \pm 1.0$ & 0.010 & 0.079 & 132 & 96 & -13.93 & 4.8 & 100 \\
     \hline
     J0926+4504 & 0.0422 & 1362.83  & $< 0.030$ & $< 2.5$ & 0.19 & 0.79 & $< 2.8$ & $< 1.3$ & -16.15 & 43.7 & - \\
     \hline
     J1132+2459 & 0.0323 & 1375.95  & $1.128 \pm 0.047$ & $55.8 \pm 2.3$ & 0.10 & 1.6 & 35 & 55 & -14.82 & 1.6 & 110 \\
     \hline
     J1146+0342 & 0.0447 & 1359.68  & $< 0.031$ & $< 2.9$ & 0.13 & 2.5 & $< 1.2$ & $< 2.2$ & -16.29 & 3.4 & - \\
     \hline
     J1149+4525 & 0.0105 & 1405.60  & $0.275 \pm 0.027$ & $1.4 \pm 0.1$ & 0.011 & 0.063 & 22 & 13 & -11.98 & 5.8 & 40 \\
     \hline
     J1201+1732 & 0.0440 & 1360.58  & $0.138 \pm 0.012$ & $12.7 \pm 1.1$ & 0.056 & 1.3 & 9.2 & 23 & -14.87 & 6.8 & 80 \\
     \hline
     J1204+3538 & 0.0339 & 1373.81  & $< 0.048$ & $< 2.7$ & 0.035 & 0.032 & $< 84$ & $< 7.7$ & -14.25 & 8.6 & - \\
     \hline
     J1220+4915 & 0.0123 & 1403.18  & $< 0.071$ & $< 0.5$ & 0.007 & 0.020 & $< 25$ & $< 7.3$ & -12.40 & 17.4 & - \\
     \hline
     J1248+4123 & 0.0269 & 1383.15  & $< 0.096$ & $< 3.4$ & 0.016 & 0.016 & $< 204$ & $< 21$ & -12.83 & 11.2 & - \\
     \hline
     J1321+4727 & 0.0147 & 1399.84  & $0.134 \pm 0.019$ & $1.4 \pm 0.2$ & 0.001 & 0.20 & 7.4 & 136 & -13.27 & 1.2 & 40 \\
     \hline
     J1359-0103 & 0.0244 & 1386.62  & $< 0.136$ & $< 3.9$ & 0.37 & 5.0 & $< 0.8$ & $< 1.0$ & -16.25 & 3.1 & - \\
     \hline
     J1359+5726 & 0.0338 & 1373.93  & $0.154 \pm 0.014$ & $8.4 \pm 0.8$ & 2.1 & 12.6 & 0.6 & 0.4 & -17.28 & 3.6 & 140 \\
     \hline
     J1402+3033 & 0.0252 & 1385.55  & $< 0.070$ & $< 2.1$ & 0.011 & 0.13 & $< 17$ & $< 19.0$ & -13.12 & 8.1 & - \\
     \hline
     J1444+0409 & 0.0388 & 1367.42  & $< 0.041$ & $< 3.0$ & 0.27 & 0.32 & $< 9.3$ & $< 1.1$ & -16.62 & 21.3 & - \\
     \hline
     J2247+0407 & 0.0378 & 1368.73  & $< 0.041$ & $< 2.9$ & 0.088 & 0.013 & $< 248$ & $< 3.3$ & -15.39 & 4.4 & - \\
     \hline
\end{tabular}
\caption{The columns are (1)~the Green Pea identifier, in J2000 coordinates, (2)~the GP redshift, (3)~the expected redshifted \hii\ line frequency, in MHz, (4)~the velocity-integrated \hii\ line flux density and error, or 3$\sigma$ upper limits to the \hii\ line flux density, in Jy~\kms, (5)~the \hi\ mass and error, or 3$\sigma$ upper limits to the \hi\ mass, in $\rm 10^8 \,  \Msun$, (6)~the SFR, in $\Msun$~yr$^{-1}$ \citep{jiang19}, (7)~the stellar mass, in $\rm 10^8 \, \Msun$ \citep{jiang19}, (8)~the \hi-to-stellar mass ratio, $\rm f_{HI} = M_{HI}/M_{*}$, (9)~the \hi\ depletion timescale, $\rm \tau_{dep} = M_{HI}/SFR$, in Gyr, (10)~the absolute B-band magnitude, $\rm M_B$, (9)~the O32 ratio, and (10)~for the \hii\ detections, the FWHM of the \hii\ line, $\rm W_{50}$, in \kms.}
\label{tab:gp-data}
\end{table}

\section{Discussion} 
\label{sec:discuss}

The overarching goal of our \hii\ studies of GPs is to examine the \hi\ conditions that result in LyC leakage in these systems. Unfortunately, at the low redshifts of the GPs of our sample, direct measurements of the LyC escape fraction are not possible today. We must hence use an indirect indicator of LyC leakage to infer the dependence of the LyC leakage on the \hi\ properties of GPs. We focus on the extinction-corrected line luminosity ratio O32~$\equiv$~[O{\sc iii}]$\lambda 5007 + \lambda 4959$/[O{\sc ii}]$\lambda$3727,3729, which has been shown to be a good indicator of LyC escape in galaxies at both low and high redshifts \citep[e.g.][]{nakajima14,izotov18,flury22}. Specifically, \citet{flury22} used the LzLCS survey to find that the LyC escape fraction correlates strongly with O32: more than half the galaxies with O32~$\gtrsim 10$ show LyC emission, while the LyC emitter fraction drops to $\lesssim 20$\% below this O32 threshold \citep[see Fig.~8 of ][]{flury22}. Most of the GPs with \hii\ searches today have measured O32 values \citep[which is not the case for other indicators of LyC leakage; e.g. the velocity separation between Ly$\alpha$ peaks, the UV slope, etc; ][]{izotov18,flury22,chisholm22,jaskot24}. We will hence compare the \hi\ properties of GPs with their O32 properties, using O32~$=10$ as the threshold above which galaxies are expected to show LyC leakage.

So far, there have been three surveys for \hii\ emission in GPs: 
\citet{kanekar21} used the Arecibo Telescope and the GBT to search for \hii\ emission from 40 GPs at $z \approx 0.02-0.09$, with 19 detections and 21 upper limits to the \hii\ line flux density . \citet{chandola24} used the Five-hundred-meter Aperture Spherical radio Telescope (FAST) to observe 28 GPs at $z \lesssim 0.05$, with 2 detections and 26 upper limits. In the present work, we obtained 7 detections and 17 non-detections of \hii\ emission in our GBT survey of GPs at $z \lesssim 0.045$. Finally,  \citet{dutta24} report a tentative detection of \hii\ emission in their Giant Metrewave Radio Telescope (GMRT) \hii\ study of J1509+3731, which was not detected by either \citet{kanekar21} or \citet{chandola24}.

We chose to restrict the full sample of GPs with \hii\ searches to redshifts $z < 0.05$, to ensure that all searches have similar sensitivity to \hii\ line emission. Excluding overlaps, we have 73 GPs at $z < 0.05$ with searches for \hii\ emission. We further excluded two GPs from the sample of \citet{chandola24} as these did not have measured O32 ratios. Finally, the current sample of GPs with \hii\ emission searches and O32~$> 10$ is limited to the stellar mass range $\rm 10^6 - 10^9 \, \Msun$. To ensure that the two O32 subsamples have similar stellar mass distributions, allowing a fair comparison of their \hi\ properties, we restricted the full GP sample to this stellar mass range. In passing, we note that using the full stellar mass range of the 71 GPs of the sample does not significantly alter our results. Our final GP sample contains 60 GPs at $z < 0.05$, 32 with O32~$< 10$ and 28 with O32~$> 10$, and with 19 detections and 41 non-detections of \hii\ emission. 

\begin{figure}[t!]
\centering
\includegraphics[width=0.9\textwidth]{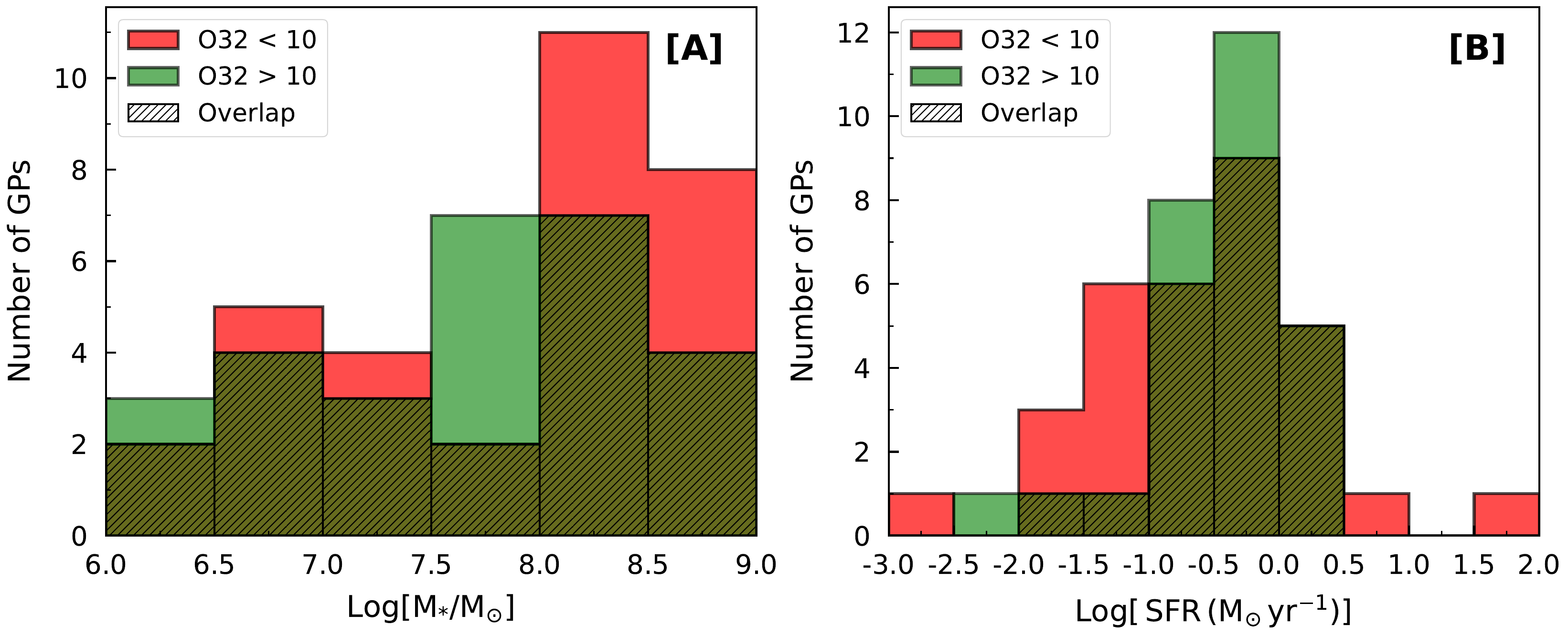}
\caption{bf Histograms showing the distributions of [A]~the stellar masses and [B]~the SFRs of the final sample of 60 GPs. In both panels, the red histogram is for GPs with O32~$< 10$ and the green one for those with O32~$> 10$, with the dashed olive region indicating the region of overlap.
\label{fig:mstar-sfr}}
\end{figure}

The O32 values, SFRs, stellar masses, and absolute B-magnitudes for the sample were obtained from the compilations of \citet{yang17} and \citet{jiang19}. {Figure~\ref{fig:mstar-sfr} shows the histograms of the distributions of the stellar masses and SFRs of the final sample of 60 GPs. The distributions for GPs with O32~$< 10$ (red) and O32~$> 10$ (green) are seen to be similar in both stellar mass and SFR.} We used a Peto-Prentice two-sample test to formally compare the stellar mass and SFR distributions of the subsamples of GPs with O32~$< 10$ and O32~$>10$. For both quantities, the distributions were found to be consistent (within $\approx 1.4\sigma$ significance) with the null hypothesis that the two subsamples are drawn from the same underlying distribution. This demonstrates that the stellar properties of the GPs of our sample show no dependence on their O32 values.

\begin{figure}[t!]
\centering
\includegraphics[width=0.75\textwidth]{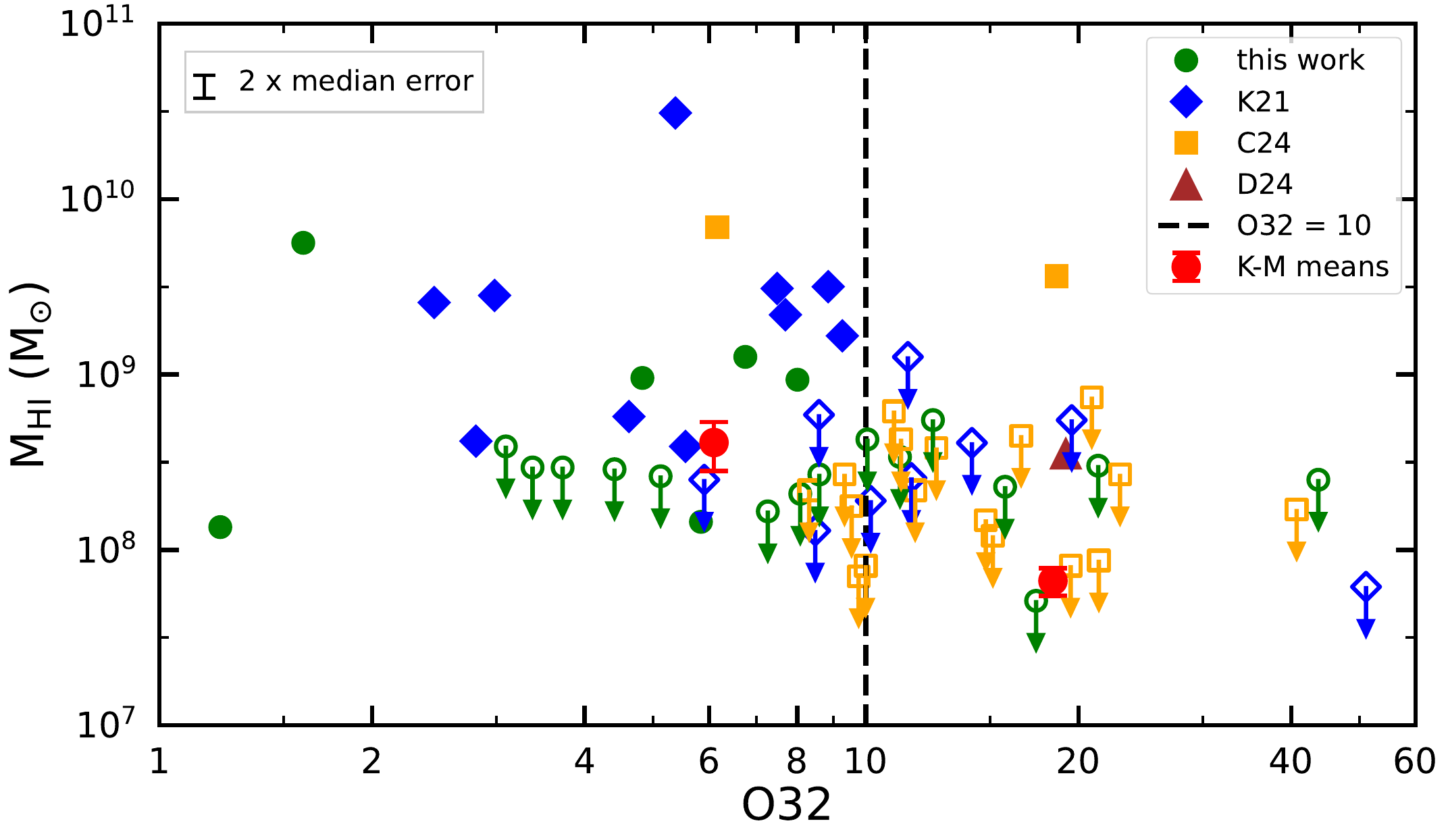}
\caption{The H{\sc i} mass, or the $3\sigma$ upper limits to the \hi\ mass, plotted against the O32 ratio for the 60 GPs of the sample. The dashed vertical line indicates the value O32~$=10$, above which galaxies are expected to show significant LyC leakage. The filled and open symbols indicate, respectively, detections and non-detections of \hii\ emission, with the green circles indicating GPs from the present work, the blue diamonds from \citet[][K21]{kanekar21}, the orange squares from \citet[][C24]{chandola24}, and the purple triangle from \citet[][D24]{dutta24}. The red circles with errors indicate the Kaplan-Meier (K-M) means of the \hi\ mass of the O32~$< 10$ and O32~$> 10$ subsamples. {The median error bar (i.e. $2 \times $~median error) for the \hi\ masses is indicated at the top left.}
\label{fig:mhi-o32}}
\end{figure}

Figure~\ref{fig:mhi-o32} plots the \hi\ masses (or $3\sigma$ upper limits on the \hi\ mass) of the 60 GPs of the sample against their O32 values. The filled symbols\footnote{The errors on the individual measurements include a 10\% error to account for systematic uncertainties in the flux density scale, which we have added in quadrature to the measurement errors.} represent detections of \hii\ emission, while the open symbols with downward-pointing arrows indicate \hii\ non-detections; see the figure caption for additional details on the symbols. 
There are far more \hii\ detections for GPs with O32~$ < 10$ than for GPs with O32~$> 10$, with 17 detections of \hii\ emission amongst the 32 GPs with O32~$< 10$, and only 2 detections \citep[including one tentative detection; ][]{dutta24} amongst the 28 GPs with O32~$> 10$. These correspond to \hii\ detection rates of $\approx 53^{+16}_{-13}$\% (O32~$< 10$) and $\approx 7.1^{+9.4}_{-4.6}$\% (O32~$> 10$), where the errors are 68\% confidence intervals from Poisson statistics \citep{gehrels86}. It is thus clear that the \hii\ detection fraction is significantly higher for GPs with O32~$< 10$ than for those with O32~$> 10$.

We next compared the distributions of the \hi\ masses of the GPs in the two O32 bins via a two-sample Peto-Prentice test, using survival analysis to take into account upper limits on the \hi\ mass \citep[as implemented in the {\sc asurv} software;][]{feigelson85,isobe86}. {Errors on the measurements of the \hi\ masses were handled through a Monte Carlo approach, using the measured values of \hi\ masses and associated errors for each GP to generate $10^4$ sets of pairs of \hi\ masses and O32 values. We ignored the measurement uncertainties on the O32 values, which are typically small relative to the $\gtrsim 10$\% uncertainty in the \hi\ mass estimates. The statistical significance of the result (quoted below) is the average of the values obtained in the $10^4$ trials.}
We obtain a probability of $2.8 \times 10^{-4}$ that the GPs of the two O32 subsamples are drawn from the same underlying distribution; in other words, we rule out the null hypothesis that the two subsamples are drawn from the same distribution at {$\approx (3.65 \pm 0.04)\sigma$ significance, where the error on the significance was obtained from the above Monte Carlo analysis.} We thus find that the distribution of \hi\ masses amongst the two GP subsamples is significantly different, with the 32 GPs with O32~$< 10$ having typically higher \hi\ masses. We further note that the Kaplan-Meier mean \hi\ mass is $\rm \langle M_{HI} \rangle = (4.1 \pm 1.3) \times 10^8 \, \Msun$ for the GPs with O32~$< 10$, $\approx 6$ times higher than the corresponding value, $\rm \langle M_{HI} \rangle = (6.7 \pm 1.1) \times 10^7 \, \Msun$, for the GPs with O32~$> 10$.

\begin{figure}
\centering
\includegraphics[width=0.75\textwidth]{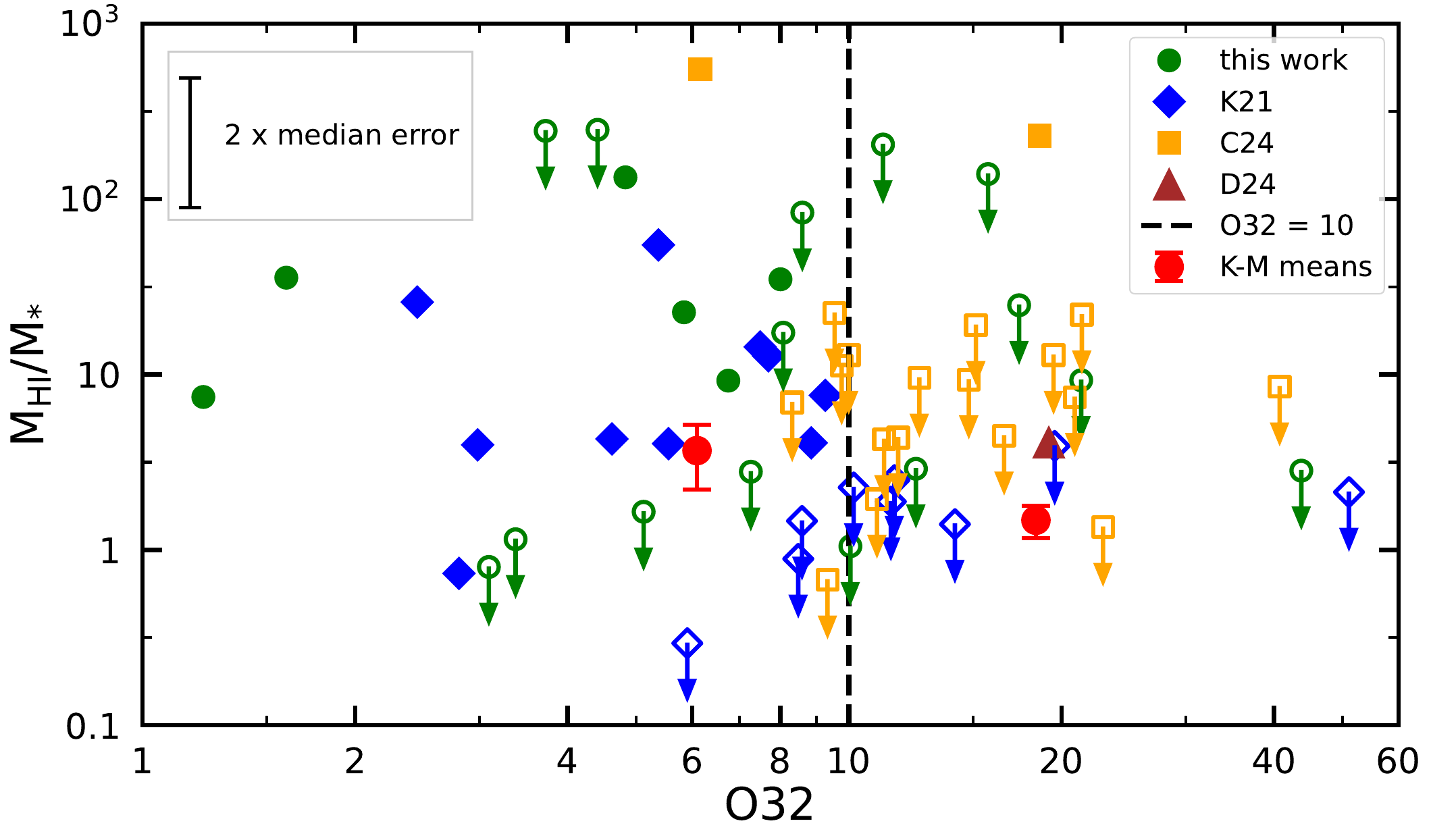}
\caption{The \hi-to-stellar mass ratio $\rm f_{HI}$, or the $3\sigma$ upper limits to $\rm f_{HI}$, plotted against the O32 ratio for the 60 GPs of the sample. The symbols are the same as those in Figure~\ref{fig:mhi-o32}. {The median error bar (i.e. $2 \times$~median error) for the \hi-to-stellar mass ratios is indicated at the top left.}
\label{fig:fhi-o32}}
\vskip -0.15in
\end{figure}

Figure~\ref{fig:fhi-o32} shows the \hi-to-stellar mass ratio $\rm f_{HI} \equiv M_{HI}/M_*$ plotted against the O32 ratio for the 60 GPs of our sample, with the filled and open symbols again indicating detections and non-detections of \hii\ emission. The figure suggests that GPs with O32~$< 10$ have higher values of $\rm f_{HI}$: the Kaplan-Meier mean value is a factor of $\approx 2.5$ higher for GPs with O32~$< 10$ than for those with O32~$> 10$. We compared the distributions of $\rm f_{HI}$ values of the two GP subsamples, again using a Peto-Prentice two-sample test and survival analysis, { and the above Monte Carlo approach to handle errors in the estimates of the \hi-to-stellar mass ratio. Here, we assumed a 0.3~dex error in the stellar mass estimates, dominated by the assumptions in the templates used to fit the spectral energy distribution.} The Peto-Prentice test yielded a probability of $1.3 \times 10^{-3}$ that the two subsamples stem from the same underlying distribution. We thus rule out the above null hypothesis at $(3.25 \pm 0.19)\sigma$ significance for the \hi-to-stellar mass ratio. 

\begin{figure}
\centering
\includegraphics[width=0.75\textwidth]{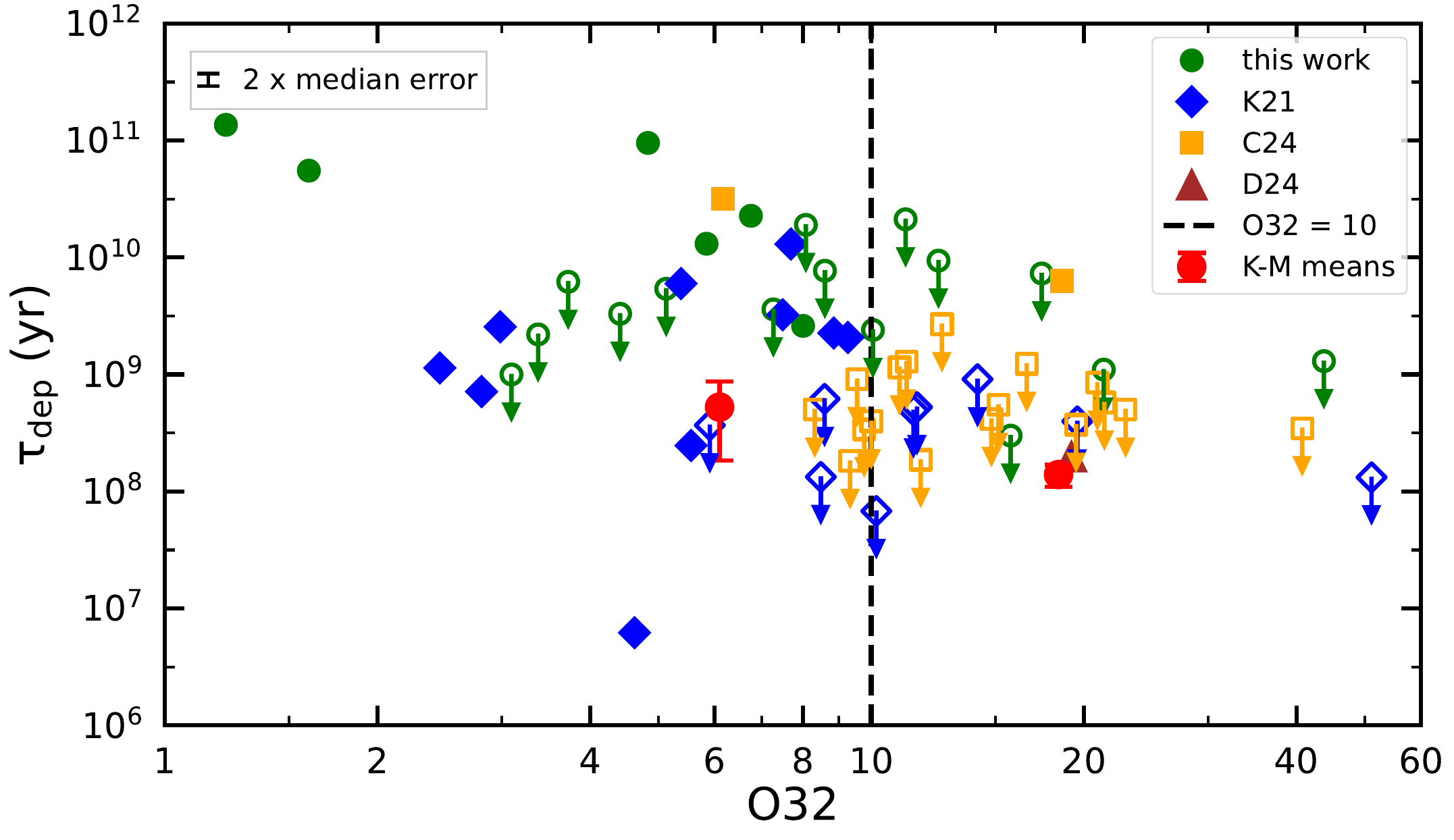}
\caption{The \hi\ depletion timescale ($\rm \tau_{dep}=M_{HI}/SFR$), in yr, plotted against the O32 ratio for the 60 GPs of the sample. The symbols are the same as those in Figure~\ref{fig:mhi-o32}. { The median error bar (i.e. $2 \times$~median error) for the \hi\ depletion timescales is indicated at the top left.}
\label{fig:tdep-o32}}
\end{figure}

Finally, Figure~\ref{fig:tdep-o32} plots the \hi\ depletion timescale $\tau_{dep} \equiv \rm M_{HI}/SFR $ against the O32 ratio. $\tau_{dep}$ gives the timescale on which the neutral atomic gas would be entirely consumed in the process of star-formation, if the \hi\ reservoir is not replenished. GPs with O32~$< 10$ are seen to have longer \hi\ depletion timescales than GPs with O32~$> 10$: the Kaplan-Meier mean of the former group is a factor of $\approx 3.8$ higher than that of the latter group. A Peto-Prentice two-sample test { (following the above Monte Carlo approach, and ignoring the errors in the SFR estimates which are small compared to the $\gtrsim 10$\% uncertainty in the \hi\ mass estimates)} finds that the probability that the two subsamples are drawn from the same underlying distribution is $\approx 2.6 \times 10^{-4}$. We thus rule out the null hypothesis that the \hi\ depletion timescales of the two O32 subsamples are drawn from the same distribution at $\approx (3.72 \pm 0.06)\sigma$ significance.

We thus find that GPs with O32~$< 10$ have higher detection rates of \hii\ emission, higher \hi\ masses, higher \hi-to-stellar mass ratios, and longer \hi\ depletion timescales than GPs with O32~$> 10$. Indeed, there are almost no detections (just 2 out of 28 systems, one of which is a tentative detection) of \hii\ emission from GPs with O32~$> 10$, and it is likely the current estimates of the Kaplan-Meier means for this subsample are actually upper limits to the true values. We emphasize that the stellar properties (e.g. the stellar mass and the SFR) of the two O32 subsamples are entirely consistent with each other, and it is only the \hi\ properties that show significant differences.

We thus find that GPs with O32~$> 10$, which are expected to show LyC leakage, have a far lower \hi\ content than their counterparts with lower O32 ratios, and also consume their \hi\ more quickly, resulting in a shorter \hi\ depletion timescale.  The paucity of atomic gas in GPs with high O32 ratios provides an immediate explanation for the LyC leakage in such galaxies.

Our findings are consistent with the suggestion that the highest O32 ratios arise where  H{\sc ii} regions are density-bounded, i.e., the H{\sc ii} region ends where the gas runs out, not where the photons run out \citep{nakajima13,nakajima14}.  Density-bounded conditions immediately imply high LyC leakage, and also a paucity of [O{\sc ii}], since there is no lower-ionization region surrounding the H{\sc ii} region.

Our results thus indicate that the galaxies that reionized the Universe at $z \gtrsim 6$ are likely to have consumed the bulk of their neutral gas during the starburst and to have low \hi\ contents. Such LyC-leaking galaxies are expected to also be strong Ly$\alpha$ emitters. We hence expect that searches for \hii\ emission from the epoch of reionization \citep[e.g.][]{hera22,kolopanis23,mertens25} should see a spatial anti-correlation between the \hii\ and Ly$\alpha$ emission signals.

Finally, we note that our results do not necessarily imply that GPs with O32~$< 10$ contain significant amounts of neutral atomic gas and that these galaxies are unlikely to be LyC leakers. The primary beams of the GBT, the Arecibo Telescope, and FAST are all quite large, with spatial resolutions of $\approx 180 - 550$~kpc at $z \approx 0.05$. This implies that some (or even all) of the observed \hii\ emission detected in the single-dish searches could arise from a companion galaxy, rather than from the target GP. This was found to be the case for both J0213+0056 at $z \approx 0.0399$ and J1148+2546 at $z \approx 0.0451$, where \citet{purkayastha22} and \citet{purkayastha24} used Jansky Very Large Array and GMRT \hii\ mapping studies to find that most of the single-dish \hii\ emission signal comes from companions that are merging with the GP. It is thus possible that the higher detection rates of \hii\ emission from GPs with O32~$< 10$ is because these galaxies are at an earlier stage of the merger, with some amount of neutral gas still present in either the GP or the merger partner, or where the ionized regions are still contained within \hi\ bubbles. Conversely, the low detection rate of \hii\ emission in GPs with O32~$> 10$ could arise because these are being observed towards the end of the merger process, when most of the gas in the GP and the merger companion has already been consumed in star-formation, and the ionized regions have merged, with little residual neutral gas surrounding the ionized regions. { A comparison of the optical images of the GPs with high and low O32 ratios, to test whether these might be at different stages of a putative merger, would be of much interest and will be carried out in a future work.}

\section{Summary}
\label{sec:summary}

We report a 193-hour GBT search for \hii\ emission from 30 GPs at $z < 0.045$, resulting in 7 detections of \hii\ emission and 17 \hii\ non-detections and upper limits on the \hi\ mass. Including GPs from the literature, we put together a sample of 60 GPs at $z < 0.05$, and with stellar masses in the range $\rm 10^6-10^9 \, \Msun$ and a wide range of O32 ratios, $\approx 1.2-55$. We use O32~$> 10$ as an indicator of LyC leakage, and compare the \hi\ properties of GPs with O32~$< 10$ and O32~$> 10$ to examine the dependence of LyC leakage on \hi\ properties. While the SFRs and stellar masses of the two O32 subsamples are consistent with their being drawn from the same underlying distributions, we find that GPs with O32~$< 10$ have systematically higher detection rates of \hii\ emission, higher \hi\ masses, and higher \hi-to-stellar mass ratios, and longer \hi\ depletion timescales, than GPs with O32~$> 10$. Indeed, only two out of 28 GPs of the sample with O32~$> 10$ show detections (one tentative) of \hii\ emission. The low \hi\ content of GPs wih O32~$> 10$ provides an immediate explanation for the escape of LyC photons from these low-$z$ analogs of the galaxies that reionized the Universe at $z \gtrsim 6$. We suggest that GPs with O32~$> 10$ may be at a later stage of the starburst process, when most of their neutral gas has been consumed by star formation. Finally, the low expected \hi\ content of the LyC-leaking galaxies that reionized the Universe suggests a spatial anti-correlation between Ly$\alpha$ emission and \hii\ emission in the $z \gtrsim 6$ Universe.

\begin{acknowledgements}

We thank an anonymous referee for a constructive report that improved this manuscript. AK and NK acknowledge support from the Department of Atomic Energy, under project 12-R\&D-TFR-5.02-0700. NK also acknowledges support from the Science and Engineering Research Board of the Department of Science and Technology via a J. C. Bose Fellowship (JCB/2023/000030). The Green Bank Observatory is a facility of the National Science Foundation operated under cooperative agreement by Associated Universities, Inc.

\end{acknowledgements}

\facilities{GBT.}

\software{{\sc gbtidl} \citep{gbtidl06}.}

\bibliography{ref.bib}
\bibliographystyle{aasjournalv7}

\end{document}